\documentclass[showpacs,showkeys,preprintnumbers]{revtex4}

\usepackage{graphicx}

\def\beq{\begin{equation}}
\def\eeq{\end{equation}}
\def\rmd{{\rm d}} 
\def\rmD{{\rm D}}

\def\rightcontract{\mathop{\hbox{\vrule width0.5pt height6pt%
  \vrule height0.5pt width6pt}}}

\def\leftcontractp#1{\mathop{\rlap{\hbox to 8pt{\hss$^{#1}$\hss}}%
  \hbox{\vrule height0.5pt width8pt \vrule width0.5pt  height6pt}}}
 
\def\rightcontractp#1{\mathop{%
  \hbox{\vrule width0.5pt height6pt \vrule height0.5pt width8pt}%
    \llap{\hbox to 8pt{\hss$^{#1}$\hss}}}}
 
\def\hook{\,\mathop{\breve{\,}}\nolimits\,}
\def\lefthook{\hook\kern-1.5pt}

\begin{document}

\title{Spin-geodesic deviations in the Kerr spacetime}

\author{D. Bini}
\affiliation{
Istituto per le Applicazioni del Calcolo ``M. Picone,'' CNR, I-00185 Rome, Italy\\
ICRA, University of Rome ``La Sapienza,'' I-00185 Rome, Italy\\
INFN, Sezione di Firenze, I--00185 Sesto Fiorentino (FI), Italy}

\author{A. Geralico}
  \affiliation{Physics Department and
ICRA, University of Rome ``La Sapienza,'' I-00185 Rome, Italy}

\begin{abstract}
The dynamics of extended spinning bodies in the Kerr spacetime is investigated in the pole-dipole particle approximation and under the assumption that the
spin-curvature force only slightly deviates the particle from a geodesic path.
The spin parameter is thus assumed to be very small and the back reaction on the spacetime geometry neglected.
This approach naturally leads to solve the Mathisson-Papapetrou-Dixon equations linearized in the spin variables as well as in the deviation vector, with the same initial conditions as for geodesic motion.
General deviations from generic geodesic motion are studied, generalizing previous results limited to the very special case of an equatorial circular geodesic as the reference path.
\end{abstract}



\maketitle

\section{Introduction}

The motion of a spinning test particle in a given gravitational background in the context of general relativity is described by the so called
Mathisson-Papapetrou model \cite{math37,papa51} completed with the Dixon-Tulczyjew \cite{tulc59,dixon64,dixon69,dixon70,dixon73,dixon74} supplementary conditions, under which the trajectory of the extended body is determined by an associated reference world line used for a multipole reduction.
Solving the resulting equations is in general a rather difficult task even for highly symmetric spacetimes, due to the complicate features of the forces associated with the non-geodesic motion.
The latter imply deviation from geodesic motion, which can be analytically discussed in very special contexts (see, e.g., Ref. \cite{mohs0}, where a plane gravitational wave is taken as the background spacetime). 
The simplest approach consists in performing purely numerical studies of the full nonlinear equations \cite{maeda,ver,hartl1,hartl2,sem99,singh}.
Otherwise one can search for analytic solutions which describe particle motion constrained along Killing trajectories, e.g. circular orbits \cite{bdfg2004,bdfgj2005,bgj2006}, also in the ultrarelativistic regime \cite{ply}.
Furthermore, Lyapunov stability in these cases has been investigated, e.g., in Refs. \cite{abrca,ryaba,svietal}.

In the present paper we follow a different approach to the problem, taking advantage of the condition of ``small spin" which is implicit in the model.
In fact, in order to neglect the particle back reaction on the background spacetime the length scale naturally associated with the spin should be very small compared to the one associated with the curvature tensor of the spacetime itself.
Introducing this smallness condition from the very beginning leads to a simplified set of linearized differential equations which can be analytically integrated at least in some special cases \cite{mashsingh,bgj_spindev}.

We investigate the deviation of the path of a spinning particle from a general geodesic in the Kerr spacetime, generalizing previous results of Ref. \cite{mashsingh}, whose analysis was limited to the case of circular geodesic motion on the equatorial plane taken as the reference path. 
The linearized set of equations in the spin variables as well as in the deviation vector components with initial conditions for geodesic motion leads to solutions for which the deviations are entirely due to the spin-curvature force.
It proves convenient to introduce an adapted orthonormal frame which is parallely propagated along the reference timelike geodesic congruence. 
In fact, the frame components of the spin orientation vector are constant and the covariant derivative of the displacement vector simply becomes an ordinary derivative with respect to the proper time along the geodesic path. 
Such a frame was found by Marck \cite{Marck1,Marck2}, who showed how to identify one leg of the associated spatial triad directly from the Killing-Yano tensor. 
Therefore, we start with a short review of Marck's construction of the parallel transported frame by evidentiating its geometrical properties.

We then consider a sample of geodesics and study the evolution of the components of the deviation vector as well as the features of the perturbed orbits. 
In the case of equatorial circular geodesic studied in Ref. \cite{mashsingh} the spinning particle position for a given spin orientation oscillates about the Keplerian orbit with (a combination of) the proper radial and vertical epicyclic frequencies.
The features of motion change significantly when non-circular and non-equatorial geodesics are considered as the reference path.

\section{Linearized MPD equations}

The set of Mathisson-Papapetrou-Dixon (MPD) equations to first order in spin is given by \cite{mashsingh,bgj_spindev}
\begin{eqnarray}
\label{papcoreqs1Iord}
ma(U)^\mu&=&-sH(U)^{\mu\rho} N_{\rho} \equiv F^{\rm (spin)}{}^\mu\,,\\
\label{papcoreqs2Iord}
s\frac{\rmD  N^{\mu}}{\rmd \tau}&=&0\,,
\end{eqnarray}
where $U$ denotes the timelike unit tangent vector to the ``center of mass line'' for the spinning particle used to perform a multipole reduction, $\tau$ is the proper time parameter along $U$, $a(U)=\nabla_UU$ is the associated 4-acceleration, 
${\mathcal H}(U)^{\mu\rho}= -[R^*]^{\mu\nu\rho\sigma}U_{\nu}U_{\sigma}$
is a spatial trace-free tensor orthogonal to $U$ referred to as the magnetic part of the Riemann tensor with respect to $U$, which in vacuum is also symmetric.
$N$ is a spacelike unit vector determining the spin orientation.
The total 4-momentum $P$ of the particle is aligned with $U$ in this limit, i.e. $P^{\mu}\approx mU^\mu$, with the mass $m$ of the particle remaining constant along the path.
The magnitude $s$ of the spin vector $S = s N$ is also a constant of motion and appears in the equations only through the specific spin angular momentum ratio $\sigma\equiv s/m$ (so that $\sigma$ has the dimensions of a length).

Consider now a pair of world lines with approximately the same initial data,
one a geodesic with 4-velocity $U_{\rm (g)}$, the other a world line of a spinning particle which deviates from the geodesic because of the combined effects of geodesic deviation and the spin-curvature coupling, with 4-velocity $U$. 
Solutions of the equation of motion to first order in the spin can then be found in the general form
\beq
\label{Uspindef}
U=U_{\rm (g)} +\sigma Y
\,.
\eeq
Moreover, the normalization condition $U\cdot U=-1$ (to first order in spin) constrains $Y$ so that
\beq
Y\cdot U_{\rm (g)}=0\,,
\eeq
synchronizing the proper times to first order, implying that $\tau$ can be used unambiguously for that single proper time parametrization of both world lines.
Furthermore, this implies that to first order in the spin, the spin orientation vector $N$ is parallely propagated along $U_{\rm (g)}$.
When the background spacetime admits a nontrivial Killing-Yano tensor $f$,
such a vector $N$ can be directly obtained from $f$, as shown by Marck \cite{Marck1,Marck2}. 

Let us introduce the coordinate displacement $\xi^\alpha$ associated with $Y^\alpha$ such that
\beq
\label{pertorb}
x^\mu (\tau)= x_{\rm (g)}^\mu (\tau_{\rm (g)})+\sigma\xi^\mu (\tau_{\rm (g)})\,,
\eeq
with
\beq
\frac{\rmd x^{\mu}}{\rmd \tau}=U^\mu\,, \qquad 
\frac{\rmd \xi^\mu}{\rmd \tau_{\rm (g)}}=Y^\mu\,.
\eeq
The 4-acceleration $a(U)$ to first order in spin turns out to be given by 
\begin{eqnarray}
\label{aU1_def}
a(U)^\mu&\simeq&\sigma\left[\frac{\rmd ^2 \xi^\mu}{\rmd \tau_{\rm (g)}^2}+2 \Gamma_{\rm (g)}{}^\mu{}_{\alpha\beta}\frac{\rmd \xi ^\alpha}{\rmd \tau_{\rm (g)}}
U_{\rm (g)}^\beta +
(\partial_\sigma\Gamma_{\rm (g)}{}^\mu{}_{\alpha\beta})\xi^\sigma U_{\rm (g)}^\alpha U_{\rm (g)}^\beta\right]\,, 
\end{eqnarray}
where $\Gamma_{\rm (g)}{}^\mu{}_{\alpha\beta}$ denote the Christoffel symbols evaluated at $x^\alpha=x^\alpha(\tau_{\rm (g)})$.
Introducing then the covariant derivative of $\xi$ along $U_{\rm (g)}$, i.e.
\beq
\frac{\rmD   \xi^\mu}{\rmd \tau_{\rm (g)}}=\frac{\rmd  \xi^\mu}{\rmd \tau_{\rm (g)}}+\Gamma_{\rm (g)}{}^\mu{}_{\alpha\beta}\xi^\alpha U_{\rm (g)}^\beta\,, 
\eeq
and taking the covariant derivative of this expression along $U_{\rm (g)}$ gives
\begin{eqnarray}
\frac{\rmD  ^2 \xi^\mu}{\rmd \tau_{\rm (g)}^2}
&=&\frac{\rmd ^2 \xi^\mu}{\rmd \tau_{\rm (g)}^2}+2\Gamma_{\rm (g)}{}^\mu{}_{\alpha\beta}\frac{\rmd  \xi^\alpha}{\rmd \tau_{\rm (g)}} U_{\rm (g)}^\beta-\left[R_{\rm (g)}{}^\mu{}_{\gamma\alpha\beta}-\partial_\alpha\Gamma_{\rm (g)}{}^\mu{}_{\gamma\beta}\right]\xi^\alpha U_{\rm (g)}^\beta U_{\rm (g)}^\gamma\,,
\end{eqnarray}
where $R_{\rm (g)}{}^\mu{}_{\gamma\alpha\beta}$ denote the Riemann tensor components evaluated at $x^\alpha=x^\alpha(\tau_{\rm (g)})$.
Therefore, Eq. (\ref{aU1_def}) becomes
\begin{eqnarray}
\label{aU1_def2}
a(U)^\mu&\simeq&\sigma\left[\frac{\rmD  ^2 \xi^\mu}{\rmd \tau_{\rm (g)}^2}+R_{\rm (g)}{}^\mu{}_{\gamma\alpha\beta}\xi^\alpha U_{\rm (g)}^\beta U_{\rm (g)}^\gamma\right]\nonumber\\
&=&\sigma\left[\frac{\rmD  ^2 \xi^\mu}{\rmd \tau_{\rm (g)}^2}+E(U_{\rm (g)})^\mu{}_{\alpha}\xi^\alpha\right]\,,
\end{eqnarray}
where $E(U_{\rm (g)})$ is the electric part of the Riemann tensor.
The equations of motion (\ref{papcoreqs1Iord}) are thus summarized by
\beq
\label{eqmoton}
\frac{\rmD  ^2 \xi^\mu}{\rmd \tau_{\rm (g)}^2}+E(U_{\rm (g)})^\mu{}_{\alpha}\xi^\alpha=-H(U_{\rm (g)})^{\mu\rho} N_{\rho}\,.
\eeq
In order to solve this set of equations it is convenient to introduce a parallel propagated frame along $U_{\rm (g)}$, implying that the covariant derivatives simply become ordinary derivatives with respect to the proper time and the components of the spin orientation vector in that frame are constant.

One may also consider a bunch of spinning particles with different spin $s$ and all surrounding a single geodesic.
Deviations from each pair are measured by 
\beq
\label{dev2}
x_1^\mu (\tau)-x_2^\mu (\tau)=(\sigma_1-\sigma_2)\xi^\mu (\tau_{\rm (g)})\,, \qquad
U_1^\mu - U_2^\mu = (\sigma_1-\sigma_2)Y^\mu\,.
\eeq

From Eq. (\ref{eqmoton}) it follows that both the displacement vector $\xi$ and the deviation vector $Y$ do not depend on the value of $s$ (or $\sigma$).
The latter instead affects the resulting particle orbits.
Typical values for astrophysical systems can be found, e.g., in Refs. \cite{mashsingh,sem99}.
For instance, for the motion of the Earth in the gravitational field of the Sun the associated length scales are $(s/m)_\oplus\approx200$ cm and $M=M_\odot\approx1.5\times10^{5}$ cm, leading to $\sigma\approx10^{-3}$.
The same order of magnitude is obtained in the case of the binary pulsar system PSR J0737-3039 as orbiting Sgr A$^*$, the supermassive ($M\simeq 10^6\ M_\odot$) black hole located at the Galactic Center, at a distance of $r\simeq 10^9$ Km \cite{lyne}. 
It consists of two close neutron stars, whose intrinsic rotations are negligible with respect to the orbital period of about $2.4$ hours. Considering then the binary system as a single object with reduced mass of about $0.7 \ M_\odot$ and intrinsic rotation equal to the orbital period, the spin parameter thus turns out to be equal to $\sigma\approx10^{-3}$.

\section{General geodesic motion in Kerr spacetime}

The Kerr metric  in standard Boyer-Lindquist coordinates is given by
\begin{eqnarray}
\rmd s^2 &=& -\left(1-\frac{2Mr}{\Sigma}\right)\rmd t^2 -\frac{4aMr}{\Sigma}\sin^2\theta\rmd t\rmd\phi+ \frac{\Sigma}{\Delta}\rmd r^2 +\Sigma\rmd \theta^2\nonumber\\
&&+\frac{(r^2+a^2)^2-\Delta a^2\sin^2\theta}{\Sigma}\sin^2 \theta \rmd \phi^2\,,
\end{eqnarray}
where $\Delta=r^2-2Mr+a^2$ and $\Sigma=r^2+a^2\cos^2\theta$; here $a$ and $M$ are the specific angular momentum and total mass of the spacetime solution. The event horizons are located at $r_\pm=M\pm\sqrt{M^2-a^2}$. 

Let us introduce the Carter family of fiducial observers, whose $4$-velocity and spatial unit vector in the Killing $2$-plane $<t, \phi>$ are given by
\begin{eqnarray}
u_{({\rm car} )}&=& \frac{r^2+a^2}{\sqrt{\Delta \Sigma}}\left(\partial_t +\frac{a}{r^2+a^2}\partial_\phi\right)\,,
\qquad 
u_{({\rm car} )}^\flat= \sqrt{\frac{\Delta}{\Sigma}}\left[-\rmd t +a\sin^2\theta \rmd\phi\right]\,,
\nonumber\\
\bar u_{({\rm car})}&=&\frac{a\sin\theta}{\sqrt{\Sigma}}\left(\partial_t +\frac{1}{a\sin^2\theta}\partial_\phi\right)\,, 
\qquad
\bar u_{({\rm car})}^\flat=\frac{\sin\theta}{\sqrt{\Sigma}}\left[-a\rmd t +(r^2+a^2)\rmd\phi\right]\,,
\end{eqnarray}
respectively.
Here $u_{({\rm car} )}$ and $\bar u_{({\rm car})}$ are future-oriented and rotating with positive angular velocity.
Other conventions give rise to sign differences which matter in the calculations developed below.
The vectors 
\beq
\label{carterframe}
e_1=e_{\hat r}=\frac1{\sqrt{g_{rr}}}\partial_r\,,\quad
e_2=e_{\hat \theta}=\frac1{\sqrt{g_{\theta \theta }}}\partial_\theta\,,\quad
e_3=\bar u_{({\rm car} )}\,,
\eeq
form an orthonormal spatial triad with dual 
\beq
\omega^1=\sqrt{g_{rr}}\rmd r\,, \quad 
\omega^2= \sqrt{g_{\theta \theta }} \rmd \theta\,, \quad
\omega^3=\bar u_{({\rm car})}^\flat\,,
\eeq
adapted to Carter observers $u_{({\rm car} )}=e_0$ (with dual $\omega^0=-u_{({\rm car} )}^\flat$).

\subsection{Timelike geodesics}

Timelike geodesics in the Kerr spacetime are given by
\begin{eqnarray}
U_{(\rm g)}^t=\frac{\rmd t_{\rm (g)}}{\rmd \tau_{\rm (g)}}&=& \frac{1}{\Sigma}\left[aB+\frac{(r^2+a^2)}{\Delta}P\right]\,,\nonumber \\
U_{(\rm g)}^r=\frac{\rmd r_{\rm (g)}}{\rmd \tau_{\rm (g)}}&=&\epsilon_r \frac{1}{\Sigma}\sqrt{R}\,,\nonumber \\
U_{(\rm g)}^\theta=\frac{\rmd \theta_{\rm (g)}}{\rmd \tau_{\rm (g)}}&=&\epsilon_\theta \frac{1}{\Sigma}\sqrt{\Theta}\,,\nonumber \\
U_{(\rm g)}^\phi=\frac{\rmd \phi_{\rm (g)}}{\rmd \tau_{\rm (g)}}&=& \frac{1}{\Sigma}\left[\frac{B}{\sin^2\theta}+\frac{a}{\Delta}P\right]\,,
\end{eqnarray}
where $\epsilon_r$ and $\epsilon_\theta$ are sign indicators, and
\begin{eqnarray}
P&=& E(r^2+a^2)-La\,,\nonumber\\
B&=& L-aE \sin^2\theta\,, \nonumber\\
R&=& P^2-\Delta (r^2+K)\,,\nonumber\\
\Theta&=&Q-\cos^2 \theta\left[a^2(1-E^2)+\frac{L^2}{\sin^2 \theta} \right]=K-a^2\cos^2\theta-\frac{B^2}{\sin^2\theta}\,,\nonumber\\
Q&=& K-(L-aE)^2\,,
\end{eqnarray}
where $E$, $L$ and $K$ are the energy, azimuthal angular momentum and generalized angular momentum (Carter constant) per unit mass of the particle.
The unit tangent vector to the timelike geodesics $U_{(\rm g)}$ has the following form with respect to the Carter frame
\begin{eqnarray}
U_{(\rm g)}&=& \frac{P}{\sqrt{\Delta \Sigma}} u_{({\rm car} )}+\frac{\epsilon_r\sqrt{R(r)}}{\sqrt{\Delta \Sigma}}e_1
+\frac{\epsilon_\theta\sqrt{\Theta(\theta)}}{\sqrt{\Sigma}}e_2+\frac{B}{\sin\theta\sqrt{\Sigma}} e_3\nonumber\\
&=& \gamma(U_{(\rm g)},u_{({\rm car} )}) [u_{({\rm car} )} +\nu(U_{(\rm g)},u_{({\rm car} )})^a e_a]\,.
\end{eqnarray}
As it is well known, a part from the common factor $1/\sqrt{\Sigma}$, the components of $U_{(\rm g)}$ with respect to Carter's frame are separated in their dependence on the coordinates $r$ and $\theta$.
This is also true for the coordinate components of $U_{(\rm g)}^\flat$
\beq
U_{(\rm g)}^\flat= -E\rmd t+\epsilon_r\frac{\sqrt{R(r)}}{\Delta}\rmd r
+\epsilon_\theta\sqrt{\Theta(\theta)}\rmd\theta+L\rmd\phi\,.
\eeq
We have then
\begin{eqnarray}
\gamma(U_{(\rm g)},u_{({\rm car} )})&\equiv& \gamma_{(\rm g)}= \frac{P}{\sqrt{\Delta \Sigma}}\,,  \nonumber\\
\nu(U_{(\rm g)},u_{({\rm car} )})^a e_a&\equiv& \nu_{(\rm g)}^a e_a
=\frac{\sqrt{\Delta }}{P}\left[  
\frac{\epsilon_r\sqrt{R(r)}}{\sqrt{\Delta }}e_1
+ \epsilon_\theta\sqrt{\Theta(\theta)} e_2+\frac{B}{\sin\theta } e_3\right]
\,.
\end{eqnarray}

We find it convenient to further decompose $\nu_{(\rm g)}$ along the radial direction and orthogonally to it, i.e. to introduce the following quantities
\begin{eqnarray}
\label{tree-vel-comp}
\nu_{(\rm g)}^\Vert&=&\nu_{(\rm g)}^1\,, \nonumber\\
\nu_{(\rm g)}^\perp&=&\nu_{(\rm g)}^3e_2 -\nu_{(\rm g)}^2e_3 =||\nu_{(\rm g)}^\perp||\, \hat \nu_{(\rm g)}^\perp\,,\nonumber\\ 
\nu_{(\rm g)}^\top &=&\nu_{(\rm g)}^2e_2 +\nu_{(\rm g)}^3e_3 =||\nu_{(\rm g)}^\top||\, \hat \nu_{(\rm g)}^\top
\,,
\end{eqnarray}
so that
\beq
\nu_{(\rm g)}=\nu_{(\rm g)}^\Vert e_1 + ||\nu_{(\rm g)}^\top|| \, \hat \nu_{(\rm g)}^\top\,,
\eeq
and with
\begin{eqnarray}
&&||\nu_{(\rm g)}^\perp||\equiv ||\nu_{(\rm g)}^\top||=\frac{\sqrt{\Delta }}{P} \sqrt{K-a^2\cos^2\theta}\,, \nonumber\\
&&\nu_{(\rm g)}^\Vert= \frac{\epsilon_r\sqrt{R(r)} }{P}\,,\qquad 
\gamma_{(\rm g)}^\Vert=\frac{P}{\sqrt{\Delta(r^2+K)}}\,.
\end{eqnarray}
It is worth to note that  $\nu_{(\rm g)}^\Vert$ has a sign in its definition.

\subsection{Killing-Yano tensor}

Consider the Killing-Yano tensor with its electromagnetic-like decomposition
\cite{foot}
\begin{eqnarray}
f&=&a\cos \theta \,\, [u_{({\rm car} )}^\flat \wedge \omega^1]+r\,\, [\omega^2\wedge \omega^3]\nonumber\\
&=& u_{({\rm car} )}^\flat \wedge {\mathcal E}(u_{({\rm car} )})+^*[u_{({\rm car} )}^\flat \wedge {\mathcal B}(u_{({\rm car} )})]\,.
\end{eqnarray}
Using the notation
\beq
\omega^{\alpha\beta}=\omega^\alpha \wedge \omega^\beta\,,
\eeq
$f$ and its dual $f^*$ can be written as follows
\beq
f=-a\cos \theta \,\, \omega^{01}+r\,\, \omega^{23}\,,\qquad 
f^*= a\cos \theta \,\, \omega^{23}+r\,\, \omega^{01}\,.
\eeq

\subsection{Parallel transported frame along $U_{\rm(g)}$}

Carter's frame is very special because of its property of making the electric and magnetic fields ${\mathcal E}(u_{({\rm car} )})$ and ${\mathcal B}(u_{({\rm car} )})$  parallel and both aligned with the radial direction
\beq
{\mathcal E}(u_{({\rm car} )})=a\cos \theta e_1\equiv {\mathcal E}e_1\,, \qquad {\mathcal B}(u_{({\rm car} )})=r e_1\equiv {\mathcal B}e_1\,,
\eeq
having introduced the more compact notation
\beq
||{\mathcal E}(u_{({\rm car} )})||=a\cos \theta  \equiv {\mathcal E}\,, \qquad ||{\mathcal B}(u_{({\rm car} )})||=r \equiv {\mathcal B} \,.
\eeq
The invariants of this field, $I_1= \frac12\,{\rm Tr}\, [f^2]$ and $I_2=\frac12\,{\rm Tr}\, [ff^*]$  are then
\beq
I_1={\mathcal B}^2-{\mathcal E}^2\,,\qquad I_2=2 {\mathcal E}{\mathcal B}\,,
\eeq
and are both nonzero, showing that the field is nonsingular.

As shown by Marck \cite{Marck1,Marck2}, a spacelike vector orthogonal to $U_{(\rm g)}$ and parallely transported along $U_{(\rm g)}$ is given by
\begin{eqnarray}
\widehat E_1 &\propto&  f \rightcontract U_{(\rm g)}=[{\mathcal E}\,\, u_{({\rm car} )}^\flat \wedge \omega^1 +{\mathcal B}\,\, \omega^2\wedge \omega^3]\rightcontract U_{(\rm g)}\nonumber\\
&=& \gamma_{(\rm g)}{\mathcal E}\,\, [\nu_{(\rm g)}^1 u_{({\rm car} )}^\flat + \omega^1]+\gamma_{(\rm g)} {\mathcal B} [ \nu_{(\rm g)}^3\omega^2-\nu_{(\rm g)}^2\omega^3]\nonumber\\
&=& \gamma_{(\rm g)}\left[{\mathcal E}\,\, \nu_{(\rm g)}^1 u_{({\rm car} )}^\flat + {\mathcal E}\,\,\omega^1+ {\mathcal B} \nu_{(\rm g)}^\perp\right]\,.
\end{eqnarray}
We have elucidated then the geometrical meaning of
$\widehat E_1$, in the sense that it represents then the unit spacelike direction of the electric part of the Killing-Yano tensor as \lq\lq measured'' by the geodesic observers
\begin{eqnarray}
{\mathcal E}(U_{(\rm g)})&=&f \rightcontract U_{(\rm g)}\nonumber\\
&=&\gamma_g P(u_{({\rm car} )},U_{(\rm g)})^{-1}[{\mathcal E}(u_{({\rm car} )})+\nu_g \times_{u_{({\rm car} )}}{\mathcal B}(u_{({\rm car} )})]\nonumber\\
&=&||{\mathcal E}(U_{(\rm g)})||\widehat E_1\,,
\end{eqnarray}
where $\times_{u_{({\rm car} )}}$ denotes the vector product in the local rest space of $u_{({\rm car} )}$ and $P(u_{({\rm car} )},U_{(\rm g)})=P(u_{({\rm car} )})P(U_{(\rm g)})$ is a composition of the two projection operators $P(u_{({\rm car} )})=g+u_{({\rm car} )}\otimes u_{({\rm car} )}$ and $P(U_{(\rm g)})=g+U_{(\rm g)}\otimes U_{(\rm g)}$.

Let us write the components of $\widehat E_1$ with respect to Carter's frame. To this aim we introduce
 the unit spatial vector (with respect to Carter's observers)
\beq
\hat \nu_S
=\frac{{\mathcal E}e_1+ {\mathcal B} \nu_{(\rm g)}^\perp}{\sqrt{{\mathcal E}^2+ {\mathcal B}^2 ||\nu_{(\rm g)}^\perp||^2}}
=\frac{{\mathcal E}e_1+ {\mathcal B} ||\nu_{(\rm g)}^\perp|| \hat \nu_{(\rm g)}^\perp}{\sqrt{{\mathcal E}^2+ {\mathcal B}^2 ||\nu_{(\rm g)}^\perp||^2}}\,,
\eeq
so that
\beq
\widehat E_1 =\gamma_S\left[\nu_S u_{({\rm car} )} + \hat \nu_S\right]\,,\qquad  
\nu_S
=\frac{{\mathcal E}\,\, \nu_{(\rm g)}^\Vert }{\sqrt{{\mathcal E}^2+ {\mathcal B}^2 ||\nu_{(\rm g)}^\perp||^2}}\,.
\eeq
A straightforward calculation shows that
\beq
||{\mathcal E}(U_{(\rm g)})||=\sqrt{K}\,,
\eeq
which can also have a simple explanation recalling the definition of the Killing tensor $K_{\mu\nu}=[f^2]_{\mu\nu}=f_{\mu\sigma}f^{\sigma}{}_\nu$. In fact 
\beq
K=K_\mu{}^\sigma U_{(\rm g)}^\mu U_{(\rm g)}{}_\sigma= f_{\mu\nu}f^{\nu\sigma}U_{(\rm g)}^\mu U_{(\rm g)}{}_\sigma={\mathcal E}(U_{(\rm g)})\cdot {\mathcal E}(U_{(\rm g)})=||{\mathcal E}(U_{(\rm g)})||^2\,.
\eeq
An equivalent representation for $\widehat E_1$ can be obtained by introducing  the following orthonormal frame
\beq
\begin{array}{lll}
&E_{\hat 0}=\gamma_{(\rm g)}^\Vert (u_{\rm (car)}+\nu_{(\rm g)}^\Vert e_1)\,, \qquad & E_{\hat 1}=\gamma_{(\rm g)}^\Vert (\nu_{(\rm g)}^\Vert u_{\rm (car)}+e_1)\,,\\
&&\\
&E_{\hat 2}=\hat\nu_{(\rm g)}^\top\,, &
E_{\hat 3}=\hat\nu_{(\rm g)}^\perp\,.
\end{array}
\eeq
We have
\beq
\widehat E_1={\mathcal E}\frac{\cosh \beta}{\sqrt{K}}E_{\hat 1} + {\mathcal B}\frac{\sinh \beta}{\sqrt{K}}E_{\hat 3}\,,
\eeq
where we have introduced the quantity $\beta$ so that
\beq
\label{betadef}
\cosh\beta=\sqrt{\frac{r^2+K}{\Sigma}}\,, \qquad
\sinh \beta=\sqrt{\frac{K-a^2\cos^2\theta}{\Sigma}}\,;
\eeq
the latter is also useful when writing $U_{(\rm g)}$, being
\beq
U_{\rm (g)}=\cosh\beta E_{\hat 0}+\sinh \beta E_{\hat 2}\,.
\eeq

Up to now we have seen that the electric part of the Killing-Yano tensor when \lq\lq measured'' by the geodesic observers plays a central role in identifying a 
unit spacelike vector (with respect to the geodesic observers) which has the nice geometric property of being parallely trasported along $U_{(\rm g)}$.
The associated magnetic part is instead given by
\begin{eqnarray}
{\mathcal B}(U_{(\rm g)})&=&f^* \rightcontract U_{(\rm g)}\nonumber\\
&=&\gamma_g P(u_{({\rm car} )},U_{(\rm g)})^{-1}[{\mathcal B}(u_{({\rm car} )})-\nu_g \times_{u_{({\rm car} )}}{\mathcal E}(u_{({\rm car} )})]\nonumber\\
&=&||{\mathcal B}(U_{(\rm g)})||\widehat E_2\,,
\end{eqnarray}
where
\beq
||{\mathcal B}(U_{(\rm g)})||=\sqrt{K+r^2-a^2\cos^2\theta}\,,
\eeq
and
\beq
\widehat E_2=-{\mathcal B}\frac{\cosh \beta}{\sqrt{K+r^2-a^2\cos^2\theta}}E_{\hat 1}+{\mathcal E}\frac{\sinh \beta}{\sqrt{K+r^2-a^2\cos^2\theta}}E_{\hat 3}\,. 
\eeq
It results
\beq
\widehat E_1 \cdot \widehat E_2=-\frac{{\mathcal E}{\mathcal B}}{\sqrt{K(K+r^2-a^2\cos^2\theta)}}\,.
\eeq
The explicit expression of $\widehat E_2$ with respect to Carter's frame follows easily
\beq
\widehat E_2= \gamma_T\,\, [\nu_T u_{\rm (car)}+\hat \nu_T]\,,
\eeq
where
\beq
\nu_T=-\frac{{\mathcal B}\nu_{(\rm g)}^\Vert}{\sqrt{{\mathcal E}^2||\nu_{(\rm g)}^\perp||^2+{\mathcal B}^2}} \,,\qquad \hat \nu_T=\frac{{\mathcal E}||\nu_{(\rm g)}^\perp|| \hat \nu_{(\rm g)}^\perp -{\mathcal B}e_1  }{\sqrt{{\mathcal E}^2||\nu_{(\rm g)}^\perp||^2+{\mathcal B}^2}}\,.
\eeq

Since ${\mathcal E}(U_{(\rm g)})$ and ${\mathcal B}(U_{(\rm g)})$ are not orthogonal, the latter cannot be used to identify an independent direction from $\widehat E_1$ in the local rest space of $U_{(\rm g)}$. To accomplish this instead one may form the Poynting vector
\beq
{\mathcal P}(U_{(\rm g)})={\mathcal E}(U_{(\rm g)})\times_{U_{(\rm g)}} {\mathcal B}(U_{(\rm g)})\,,
\eeq 
or simply apply a Gram-Schmidt orthogonalization procedure to $\widehat E_2$ (which is already orthogonal to $U_{(\rm g)}$).
In the latter case one immediately finds 
\begin{eqnarray}
\widehat E_2^\perp &=& \frac{\widehat E_2-(\widehat E_1 \cdot \widehat E_2)\widehat E_1}{1-(\widehat E_1 \cdot \widehat E_2)^2}\nonumber\\
&=& -\frac{{\mathcal B}\sinh \beta}{\sqrt{K}}E_{\hat 1}+\frac{{\mathcal E}\cosh\beta}{\sqrt{K}}E_{\hat 3}\,.
\end{eqnarray}
Alternatively, one may evaluate the Poynting vector, that is
\begin{eqnarray}
{\mathcal P}(U_{(\rm g)})&=& ||{\mathcal E}(U_{(\rm g)})||\, ||{\mathcal B}(U_{(\rm g)}) ||\,\, \widehat E_1 \times_{U_{(\rm g)}} \widehat E_2 \nonumber\\
&=& \sqrt{K(K+r^2-a^2\cos^2\theta)}\,\, \widehat E_1 \times_{U_{(\rm g)}} \widehat E_2 \nonumber \\
&=& ({\mathcal E}^2+{\mathcal B}^2)\sinh \beta \cosh \beta\,\,  E_{\hat 1}\times_{U_{(\rm g)}} E_{\hat 3}\nonumber\\
&=& \sqrt{(r^2+K)(K-a^2\cos^2\theta)}\,\,  E_{\hat 1}\times_{U_{(\rm g)}} E_{\hat 3}\,,
\end{eqnarray}
where the vector product is taken in the local rest space of $U_{(\rm g)}$, i.e.
\beq
[E_{\hat 1}\times_{U_{(\rm g)}} E_{\hat 3}]^\alpha= U_{(\rm g)}{}_{\mu}\eta^{\mu\alpha\beta\gamma}[E_{\hat 1}]_\beta [E_{\hat 3}]_\gamma
= (\cosh \beta [E_{\hat 0}]_\mu +\sinh \beta [E_{\hat 2}]_\mu)\eta^{\mu\alpha}{}_{\hat 1\hat 3}
=- \bar U_{\rm(g)}^\alpha\,,
\eeq
and we have used the spacetime orientation condition $\eta_{\hat 0 \hat 1 \hat 2 \hat 3}=1$.

Re-arranging signs conveniently, an adapted triad to $U_{\rm(g)}$ can be formed then as follows
\begin{eqnarray}
F_{\hat 1}&=&\bar U_{\rm(g)}=\sinh \beta E_{\hat 0}+\cosh \beta E_{\hat 2}\,,\nonumber\\
F_{\hat 2}&=&\frac{{\mathcal E}\cosh \beta}{\sqrt{K}}E_{\hat 1}+\frac{{\mathcal B}\sinh \beta}{\sqrt{K}}E_{\hat 3}=\widehat E_1\,,\nonumber\\
F_{\hat 3}&=&\frac{{\mathcal B}\sinh \beta}{\sqrt{K}}E_{\hat 1}-\frac{{\mathcal E}\cosh \beta}{\sqrt{K}}E_{\hat 3}=-\widehat E_2^\perp \,.
\end{eqnarray}
Finally, from this a parallel transported one along $U_{\rm(g)}$ results in
\begin{eqnarray}
\label{parframe}
\lambda_{\hat 1}&=&\cos\psi F_{\hat 3}-\sin\psi F_{\hat 1}\,,\nonumber\\
\lambda_{\hat 2}&=&F_{\hat 2}\,,\nonumber\\
\lambda_{\hat 3}&=&\sin\psi F_{\hat 3}+\cos\psi F_{\hat 1}\,,
\end{eqnarray}
where
\beq
\label{eqpsi}
\frac{\rmd \psi}{\rmd \tau_{\rm (g)}}=\frac{\sqrt{K}}{\Sigma}\left(\frac{P}{r^2+K}+\frac{aB}{K-a^2\cos^2\theta}\right)\,,
\eeq
which in general can be integrated in terms of elliptic functions.

\section{Deviations from geodesics}

Let us decompose the displacement $\xi$ as well as the associated deviation vector $Y$ with respect to the parallel transported frame (\ref{parframe}), i.e.
\beq
\xi=\xi^{\hat a}\lambda_{\hat a}\,, \qquad
Y=Y^{\hat a}\lambda_{\hat a}\,, \qquad
Y^{\hat a}=\frac{\rmd \xi^{\hat a}}{\rmd \tau_{\rm (g)}}\,.
\eeq
The equations of motion (\ref{eqmoton}) thus become
\beq
\label{eqmoton2}
\frac{\rmd ^2 \xi^{\hat a}}{\rmd \tau_{\rm (g)}^2}+E(U_{\rm (g)})^{\hat a}{}_{\hat b}\xi^{\hat b}=-H(U_{\rm (g)})^{\hat a}{}_{\hat b} N^{\hat b}\,.
\eeq
The electric and magnetic part of the Riemann tensor in the parallel propagated frame are listed in Appendix \ref{app1}.
These equations together with Eq. (\ref{eqpsi}) have to be integrated with initial conditions $\xi^{\hat a}(0)=0=\frac{\rmd \xi^{\hat a}}{\rmd \tau_{\rm (g)}}(0)$ and $\psi(0)=0$ and specifying the fixed spin orientation with respect to that frame, once a reference geodesic has been selected.

The perturbed orbit is given by Eq. (\ref{pertorb}), i.e.
\beq
\label{pertorb2}
t=t_{\rm (g)}+t_\sigma\,, \quad
r=r_{\rm (g)}+r_\sigma\,, \quad
\theta=\theta_{\rm (g)}+\theta_\sigma\,, \quad
\phi=\phi_{\rm (g)}+\phi_\sigma\,, 
\eeq
with
\begin{eqnarray}
t_\sigma&=&\sigma\left\{
\left[\frac{\alpha r(r^2+a^2)}{\Delta\sqrt{K}}U_{\rm (g)}^r+\frac{a^2\cos\theta\sin\theta}{\alpha\sqrt{K}}U_{\rm (g)}^\theta\right]\zeta^{\hat 1}
+\left[\frac{a\cos\theta(r^2+a^2)}{\Delta\sqrt{K}}U_{\rm (g)}^r
-\frac{ar\sin\theta}{\sqrt{K}}U_{\rm (g)}^\theta\right]\zeta^{\hat 2}\right.\nonumber\\
&&\left.
+\left[\frac{\alpha P(r^2+a^2)}{\Delta\Sigma}+\frac{aB}{\alpha\Sigma}\right]\zeta^{\hat 3}
\right\}\,, \nonumber\\
r_\sigma&=&\sigma\left\{
\frac{\alpha rP}{\Sigma\sqrt{K}}\zeta^{\hat 1}
+\frac{a\cos\theta P}{\Sigma\sqrt{K}}\zeta^{\hat 2}
+\alpha U_{\rm (g)}^r\zeta^{\hat 3}
\right\}\,, \nonumber\\
\theta_\sigma&=&\sigma\left\{
-\frac{aB\cos\theta}{\alpha\Sigma\sqrt{K}\sin\theta}\zeta^{\hat 1}
+\frac{rB}{\Sigma\sqrt{K}\sin\theta}\zeta^{\hat 2}
+\frac1{\alpha}U_{\rm (g)}^\theta\zeta^{\hat 3}
\right\}\,, \nonumber\\
\phi_\sigma&=&\sigma\left\{
\left[\frac{\alpha ar}{\Delta\sqrt{K}}U_{\rm (g)}^r+\frac{a\cos\theta}{\alpha\sqrt{K}\sin\theta}U_{\rm (g)}^\theta\right]\zeta^{\hat 1}
+\left[\frac{a^2\cos\theta}{\Delta\sqrt{K}}U_{\rm (g)}^r-\frac{r}{\sqrt{K}\sin\theta}U_{\rm (g)}^\theta\right]\zeta^{\hat 2}\right.\nonumber\\
&&\left.
+\left[\frac{\alpha aP}{\Delta\Sigma}+\frac{B}{\alpha\Sigma\sin^2\theta}\right]\zeta^{\hat 3}
\right\}\,, 
\end{eqnarray}
where $\alpha\equiv\tanh\beta$ (see Eq. (\ref{betadef})) and the following rotation about the vertical axis by an angle $\psi$ has been performed for convenience, i.e. $\zeta^{\hat 2}=\xi^{\hat 2}$ and
\beq
\label{zetadef}
\zeta^{\hat 1}=\xi^{\hat 1}\cos\psi+\xi^{\hat 3}\sin\psi\,,\qquad
\zeta^{\hat 3}=-\xi^{\hat 1}\sin\psi+\xi^{\hat 3}\cos\psi\,.
\eeq

We then consider different choices of the reference geodesic and study the corresponding deviations.

\subsection{Deviations from equatorial geodesics}

In the case of equatorial geodesics we have $\theta=\pi/2$ and $\rmd\theta/\rmd\tau_{\rm (g)}=0$, implying that
\beq
\Theta=0\,, \quad
K=(L-aE)^2\,.
\eeq
The nonvanishing components of the electric and magnetic part of the Riemann tensor in the parallel propagated frame are given by
\begin{eqnarray}
E(U_{\rm (g)})_{\hat 1\hat 1}&=&\frac{M}{r^3}+\frac{3M}{r^5}(r^2+K)\cos^2\psi\,, \nonumber\\
E(U_{\rm (g)})_{\hat 1\hat 3}&=&-\frac{3M}{r^5}(r^2+K)\cos\psi\sin\psi\,, \nonumber\\
E(U_{\rm (g)})_{\hat 2\hat 2}&=&\frac{M}{r^3}+\frac{3MK}{r^5}\,,\nonumber\\
E(U_{\rm (g)})_{\hat 3\hat 3}&=&-\frac{2M}{r^3}-\frac{3MK}{r^5}+\frac{3M}{r^5}(r^2+K)\cos^2\psi\,, 
\end{eqnarray}
and
\begin{eqnarray}
H(U_{\rm (g)})_{\hat 1\hat 2}&=&\frac{3M}{r^5}\sqrt{K(r^2+K)}\cos\psi\,, \nonumber\\
H(U_{\rm (g)})_{\hat 2\hat 3}&=&\frac{3M}{r^5}\sqrt{K(r^2+K)}\sin\psi\,,
\end{eqnarray}
respectively.
Note that now $r$ is a function of the proper time according to
\beq
\frac{\rmd r_{\rm (g)}}{\rmd \tau_{\rm (g)}}=\epsilon_r \frac{\sqrt{R}}{r^2}\,,
\eeq
and 
\beq
\frac{\rmd \psi}{\rmd \tau_{\rm (g)}}=\frac{LE+a(1-E^2)}{r^2+K}\,.
\eeq
The system (\ref{eqmoton2}) writes as
\begin{eqnarray}
\label{eqsequat}
\frac{\rmd ^2 \xi^{\hat 1}}{\rmd \tau_{\rm (g)}^2}&=&-E(U_{\rm (g)})_{\hat 1\hat 1}\xi^{\hat 1}-E(U_{\rm (g)})_{\hat 1\hat 3}\xi^{\hat 3}-H(U_{\rm (g)})_{\hat 1\hat 2}N^{\hat 2}\,, \nonumber\\
\frac{\rmd ^2 \xi^{\hat 2}}{\rmd \tau_{\rm (g)}^2}&=&-E(U_{\rm (g)})_{\hat 2\hat 2}\xi^{\hat 2}-H(U_{\rm (g)})_{\hat 1\hat 2}N^{\hat 1}-H(U_{\rm (g)})_{\hat 1\hat 3}N^{\hat 3}\,, \nonumber\\
\frac{\rmd ^2 \xi^{\hat 3}}{\rmd \tau_{\rm (g)}^2}&=&-E(U_{\rm (g)})_{\hat 1\hat 3}\xi^{\hat 1}-E(U_{\rm (g)})_{\hat 3\hat 3}\xi^{\hat 3}-H(U_{\rm (g)})_{\hat 2\hat 3}N^{\hat 2}\,.
\end{eqnarray}
If the spin is aligned along the rotation axis, i.e. $N^{\hat 1}=0=N^{\hat 3}$ and $N^{\hat 2}=-1$, we have $\xi^{\hat 2}\equiv0$.

The behavior of the frame components of the deviation vector $Y$, the magnitude of the displacement vector $\xi$ as well as the perturbed orbits corresponding to selected equatorial geodesics are shown in Figs. \ref{fig:1}--\ref{fig:4}.

If the geodesic is also circular at a given $r=$ const we have in addition $R=0=\rmd R/\rmd r$, so that
\beq
E=\frac{r^2-2Mr\pm a\sqrt{Mr}}{r\sqrt{r^2-3Mr\pm2a\sqrt{Mr}}}\,, \quad
L=\pm\frac{\sqrt{Mr}(r^2\mp2a\sqrt{Mr}+a^2)}{r\sqrt{r^2-3Mr\pm2a\sqrt{Mr}}}\,,
\eeq
where upper/lower signs correspond to co/counter-revolving orbits. 
This is the only case which can be completely solved in closed analytical form \cite{mashsingh}.
It turns out that the spinning particle position for a given spin orientation oscillates about the Keplerian orbit with (a combination of) the proper radial and vertical epicyclic frequencies, while the azimuthal motion undergoes similar oscillations plus an additional secular drift (see Fig. \ref{fig:1}).

\subsection{Deviations from spherical geodesics}

Spherical geodesics are characterized by $r=$ const and $R=0=\rmd R/\rmd r$, so that
\begin{eqnarray}
E&=&\frac{(r-M)(r^2+K)+r\Delta}{2r\sqrt{\Delta(r^2+K)}}\,, \nonumber\\
L&=&\frac{1}{2ar\sqrt{\Delta(r^2+K)}}
\{r(r^2+a^2)\Delta+(r^2+K)[M(r^2-a^2)-r\Delta]\}\,,
\end{eqnarray}
with arbitrary $K$.
The behavior of the frame components of the deviation vector $Y$, the magnitude of the displacement vector $\xi$ as well as the perturbed orbit corresponding to a selected spherical geodesic is shown in Fig. \ref{fig:5}.

Polar geodesics have in addition $L=0$, which condition fixes $K$ too
\begin{eqnarray}
E&=&\Delta\left[\frac{r}{r^2+a^2}\frac{1}{r\Delta-M(r^2-a^2)}\right]^{1/2}\,, \nonumber\\
K&=&r\frac{Mr(r^2-a^2)+a^2\Delta}{r\Delta-M(r^2-a^2)}\,.
\end{eqnarray}
We consider below the limiting case of Schwarzschild spacetime, which can be treated analytically.

\subsubsection{Deviations from polar geodesics in Schwarzschild}

In the Schwarzschild case ($a=0$) we have
\beq
E=\frac{r-2M}{\sqrt{r(r-3M)}}\,, \qquad
K=\frac{Mr^2}{r-3M}\,.
\eeq

The solution of the geodesic equations is given by
\beq
\label{geopolschw}
t_{\rm (g)}=\Gamma_K\,\tau_{\rm (g)}+t_0\,, \quad
r_{\rm (g)}=r_0\,, \quad
\theta_{\rm (g)}=\epsilon_\theta\eta_K\tau_{\rm (g)}+\theta_0\,, \quad
\phi_{\rm (g)}=\phi_0\,, 
\eeq
where
\beq
\Gamma_K=\sqrt{\frac{r_0}{r_0-3M}}\,, \qquad
\eta_K=\frac{1}{r_0}\sqrt{\frac{M}{r_0-3M}}\,,
\eeq
the latter being the well known orbital frequency governing the geodesic oscillations out of the equatorial plane in the case of equatorial circular geodesics, so that
\beq
\label{psipolschw}
\psi=\epsilon_\theta\zeta_K\tau_{\rm (g)}\,, \qquad 
\zeta_K=\sqrt{\frac{M}{r_0^3}}\,.
\eeq

The system (\ref{eqmoton2}) in terms of the variable $\psi$ writes as
\begin{eqnarray}
\label{eqschwpolar}
\frac{\rmd ^2 \xi^{\hat 1}}{\rmd \psi^2}&=&(3\gamma_K^2\cos^2\psi-1)\xi^{\hat 1}+3\gamma_K^2\cos\psi\sin\psi\xi^{\hat 3}-3\gamma_K^2\nu_K N^{\hat 2}\cos\psi\,, \nonumber\\
\frac{\rmd ^2 \xi^{\hat 2}}{\rmd \psi^2}&=&-\Gamma_K^2\xi^{\hat 2}-3\gamma_K^2\nu_K (N^{\hat 1}\cos\psi+N^{\hat 3}\sin\psi)\,, \nonumber\\
\frac{\rmd ^2 \xi^{\hat 3}}{\rmd \psi^2}&=&3\gamma_K^2\cos\psi\sin\psi\xi^{\hat 1}-(3\gamma_K^2\cos^2\psi-1-\Gamma_K^2)\xi^{\hat 3}-3\gamma_K^2\nu_K N^{\hat 2}\sin\psi\,,
\end{eqnarray}
with initial conditions $\xi^{\hat a}=0=\frac{\rmd \xi^{\hat a}}{\rmd \psi}$ at $\psi=0$.
Equations (\ref{eqschwpolar}) reduce to an autonomous system by the transformation  (\ref{zetadef}), so that the rotated coordinate axes correspond to the radial, vertical, and tangential directions, i.e.
\begin{eqnarray}
\label{eqschwpolar2}
\frac{\rmd ^2 \zeta^{\hat 1}}{\rmd \psi^2}&=&3\gamma_K^2\zeta^{\hat 1}+2\frac{\rmd \zeta^{\hat 3}}{\rmd \psi}-3\gamma_K^2\nu_K N^{\hat 2}\,, \nonumber\\
\frac{\rmd ^2 \zeta^{\hat 2}}{\rmd \psi^2}&=&-\Gamma_K^2\zeta^{\hat 2}-3\gamma_K^2\nu_K (N^{\hat 1}\cos\psi+N^{\hat 3}\sin\psi)\,, \nonumber\\
\frac{\rmd ^2 \zeta^{\hat 3}}{\rmd \psi^2}&=&-2\frac{\rmd \zeta^{\hat 1}}{\rmd \psi}\,,
\end{eqnarray}
with initial conditions $\zeta^{\hat a}=0=\frac{\rmd \zeta^{\hat a}}{\rmd \psi}$ at $\psi=0$.
The solution is 
\begin{eqnarray}
\label{solzetaa}
\zeta^{\hat 1}&=&\frac{3\gamma_K^2\nu_K}{\rho_K^2} N^{\hat 2}(\cos\rho_K\psi-1)\,, \nonumber\\
\zeta^{\hat 2}&=&\frac1{\nu_K}\left[(\cos\Gamma_K\psi-\cos\psi)N^{\hat 1}+\frac1{\Gamma_K}(\sin\Gamma_K\psi-\Gamma_K\sin\psi)N^{\hat 3}\right]\,, \nonumber\\
\zeta^{\hat 3}&=&-\frac{6\gamma_K^2\nu_K}{\rho_K^3} N^{\hat 2}(\sin\rho_K\psi-\rho_K\psi)\,,
\end{eqnarray}
where
\beq
\rho_K=\sqrt{4-3\gamma_K^2}=\sqrt{\frac{r_0-6M}{r_0 -3M}}\,,
\eeq
so that $\Omega_{\rm(ep)}\equiv\rho_K\zeta_K$ is the well known epicyclic frequency governing the radial perturbations of equatorial circular geodesics.

Finally, the perturbations to the polar geodesic (\ref{geopolschw}) due to spin are given by
\begin{eqnarray}
t_\sigma&=&\sigma\Gamma_K\nu_K\zeta^{\hat 3}\,, \nonumber\\
r_\sigma&=&\sigma\frac{r_0\zeta_K}{\nu_K}\zeta^{\hat 1}\,, \nonumber\\
\theta_\sigma&=&\sigma\epsilon_\theta\frac{\gamma_K}{r_0}\zeta^{\hat 3}\,, \nonumber\\
\phi_\sigma&=&-\sigma\epsilon_\theta\frac{1}{r_0\sin\Gamma_K\psi}\zeta^{\hat 2}\,, 
\end{eqnarray}
which can be also expressed in terms of the variables $\theta_{\rm (g)}$ or $\tau_{\rm (g)}$ through Eqs. (\ref{geopolschw}) and (\ref{psipolschw}), respectively.

Consider a spinless particle moving along a polar geodesic starting from the axis, i.e. the orbit (\ref{geopolschw}) with $\theta_0=0$. The path will be confined on a plane at constant radius and constant value of the azimuthal angle.
If the particle is endowed with spin, instead, the orbit no longer will maintain spherical and in addition will be dragged in the sense of increasing/decreasing values of the azimuthal angle depending on the spin orientation. After a quarter of revolution, i.e. for $\theta_{\rm (g)}$ from $0$ to $\pi/2$, the radial shift and node advance when the orbit crosses the equatorial plane are given by
\begin{eqnarray}
\delta r\equiv r_s(\theta_{\rm (g)}=\pi/2)&=&\sigma\frac{3r_0\zeta_K\gamma_K^2}{\rho_K^2} N^{\hat 2}\left[\cos\left(\frac{\rho_K}{\Gamma_K}\frac{\pi}{2}\right)-1\right]\,, \nonumber\\
\delta \phi\equiv\phi_s(\theta_{\rm (g)}=\pi/2)&=&\sigma\frac1{r_0\nu_K}\left\{\cos\left(\frac{1}{\Gamma_K}\frac{\pi}{2}\right)N^{\hat 1}-\left[\frac1{\Gamma_K}-\sin\left(\frac{1}{\Gamma_K}\frac{\pi}{2}\right)\right]N^{\hat 3}\right\}\,. 
\end{eqnarray}
In the special case $N=-\lambda_{\hat 2}$, $\delta\phi=0$ and
\beq
\delta r=-\sigma\frac{3r_0\zeta_K\gamma_K^2}{\rho_K^2} \left[\cos\left(\frac{\rho_K}{\Gamma_K}\frac{\pi}{2}\right)-1\right]\,,
\eeq
which monotonically decreases with increasing $r_0$ from its maximum value 
\beq
\delta r_{\rm max}=\frac1{\sqrt{6}}\left(\frac{\pi}{2}\right)^2\sigma\
\eeq
attained at $r_0=6M$.

Moreover, the deviations (\ref{dev2}) between a pair of such particles are 
\beq
||x_1-x_2||=|\sigma_1-\sigma_2|\cdot||\xi||\,,
\eeq
where $||\xi||=||\zeta||$ depends on the spin orientation, the proper time parameter and the (constant) radius of the reference orbit.
When $N^{\hat 2}\not=0$, the deviation linearly increases with $\psi$ due to the presence of a secular term in $\zeta^{\hat 3}$.
If instead $N^{\hat 2}=0$, the deviations are oscillations about the reference path.

\subsection{Deviations from geodesics along the symmetry axis}

The motion along the symmetry axis requires $L=0$ and $\theta=0$, so that $K=a^2$.
The nonvanishing components of the electric and magnetic part of the Riemann tensor in the parallel propagated frame are given by \cite{punslychmash}
\beq
E(U_{\rm (g)})_{\hat 1\hat 1}=-\frac{Mr(3a^2-r^2)}{(r^2+a^2)^3}=E(U_{\rm (g)})_{\hat 3\hat 3}=-\frac12E(U_{\rm (g)})_{\hat 2\hat 2}\,,  
\eeq
and
\beq
H(U_{\rm (g)})_{\hat 1\hat 1}=-\frac{aM(a^2-3r^2)}{(r^2+a^2)^3}=H(U_{\rm (g)})_{\hat 3\hat 3}=-\frac12H(U_{\rm (g)})_{\hat 2\hat 2}\,,
\eeq
respectively.
Note that $r$ is a function of the proper time according to
\beq
\frac{\rmd r_{\rm (g)}}{\rmd \tau_{\rm (g)}}
=\epsilon_r \frac{\sqrt{R}}{r^2+a^2}
=\epsilon_r \sqrt{E^2-1+\frac{2Mr}{r^2+a^2}}\,,
\eeq
and 
\beq
\frac{\rmd \psi}{\rmd \tau_{\rm (g)}}=\frac{aE}{r^2+a^2}\,.
\eeq
The system (\ref{eqmoton2}) in terms of the variable $r$ writes as
\begin{eqnarray}
\label{eqsaxis}
R(r)\frac{\rmd ^2 \xi^{\hat 1}}{\rmd r^2}&=&
M(r^2-a^2)\frac{\rmd \xi^{\hat 1}}{\rmd r}+Mr\frac{3a^2-r^2}{r^2+a^2}\xi^{\hat 1}+aM\frac{a^2-3r^2}{r^2+a^2}N^{\hat 1}\,, \nonumber\\
R(r)\frac{\rmd ^2 \xi^{\hat 2}}{\rmd r^2}&=&
M(r^2-a^2)\frac{\rmd \xi^{\hat 2}}{\rmd r}-2Mr\frac{3a^2-r^2}{r^2+a^2}\xi^{\hat 2}-2aM\frac{a^2-3r^2}{r^2+a^2}N^{\hat 2}\,, \nonumber\\
R(r)\frac{\rmd ^2 \xi^{\hat 3}}{\rmd r^2}&=&
M(r^2-a^2)\frac{\rmd \xi^{\hat 3}}{\rmd r}+Mr\frac{3a^2-r^2}{r^2+a^2}\xi^{\hat 3}+aM\frac{a^2-3r^2}{r^2+a^2}N^{\hat 3}\,,
\end{eqnarray}
with initial conditions $\xi^{\hat a}=0=\frac{\rmd \xi^{\hat a}}{\rmd r}$ at $r=r_0$.
If the spin is aligned along the axis, i.e. $N^{\hat 1}=0=N^{\hat 3}$ and $N^{\hat 2}=-1$, we have $\xi^{\hat 1}\equiv0\equiv\xi^{\hat 3}$ and
\beq
\xi^{\hat 2}=2aM\left(E^2-1+\frac{2Mr}{r^2+a^2}\right)^{1/2}\left[
W(r)\left(\frac{r}{(r^2+a^2)^2}-\frac{r_0}{(r_0^2+a^2)^2}\right)-\int_{r_0}^rW(r)\frac{a^2-3r^2}{r^2+a^2}\rmd r
\right]\,,
\eeq
where
\beq
W(r)=\int^r\left(E^2-1+\frac{2Mr}{r^2+a^2}\right)^{-3/2}\rmd r\,,
\eeq
which can be expressed in terms of elliptic functions.
In this case, the perturbations to the reference geodesic due to spin are then given by
\begin{eqnarray}
t_\sigma&=&\sigma\epsilon_r\frac{\sqrt{R(r)}}{\Delta}\xi^{\hat 2}\,, \nonumber\\
r_\sigma&=&\sigma E\xi^{\hat 2}\,, \nonumber\\
\theta_\sigma&=&0\,, \nonumber\\
\phi_\sigma&=&-\sigma\epsilon_r\frac{a}{r^2+a^2}\frac{\sqrt{R(r)}}{\Delta}\xi^{\hat 2}\,, 
\end{eqnarray}
where the result $\theta_\sigma=0$ is consistent with the choice of spin aligned with the rotation axis.

% figure 1

\begin{figure} 
\typeout{*** EPS figure 1}
\begin{center}
$\begin{array}{cc}
\includegraphics[scale=0.25]{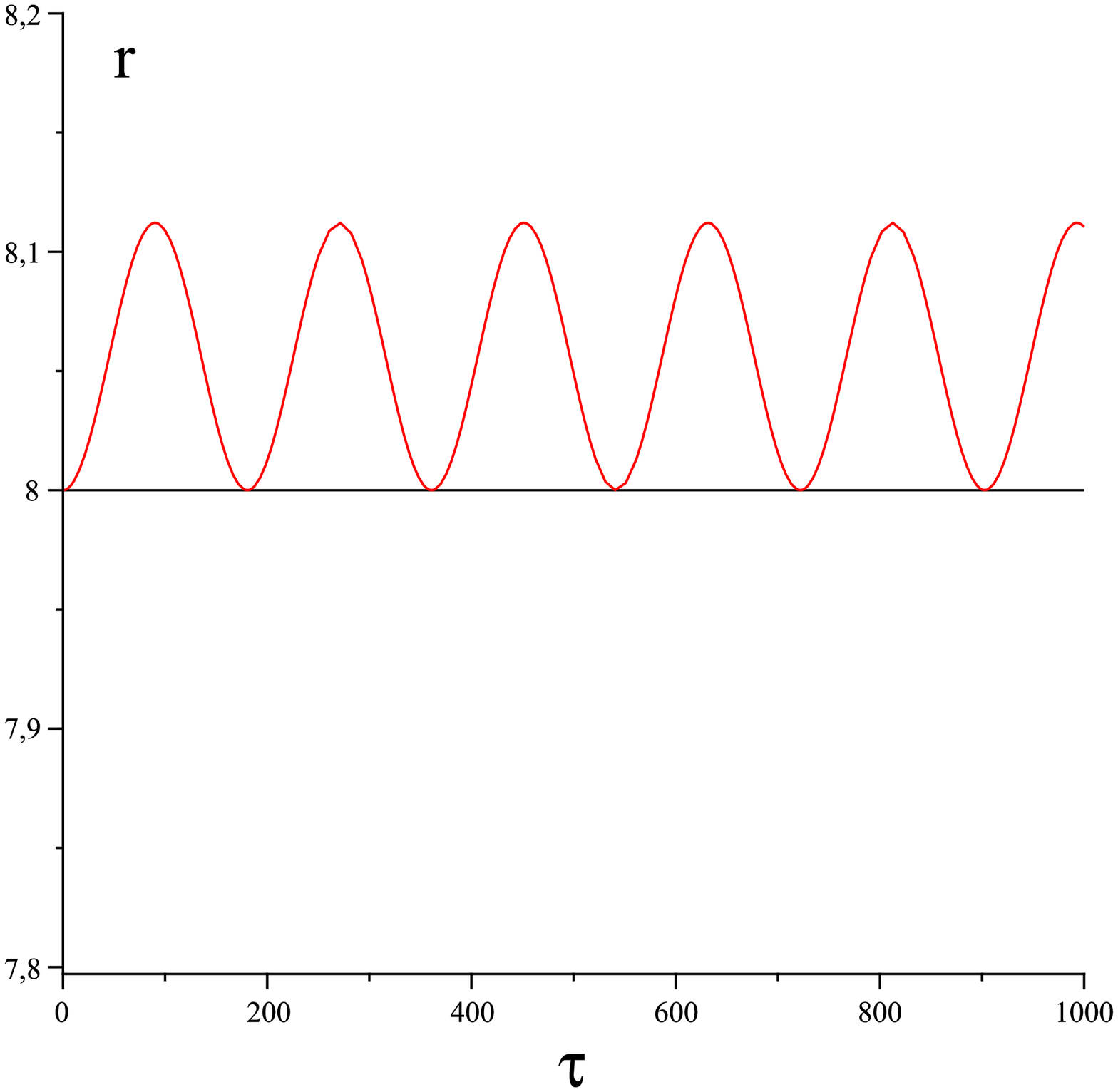}&\quad
\includegraphics[scale=0.25]{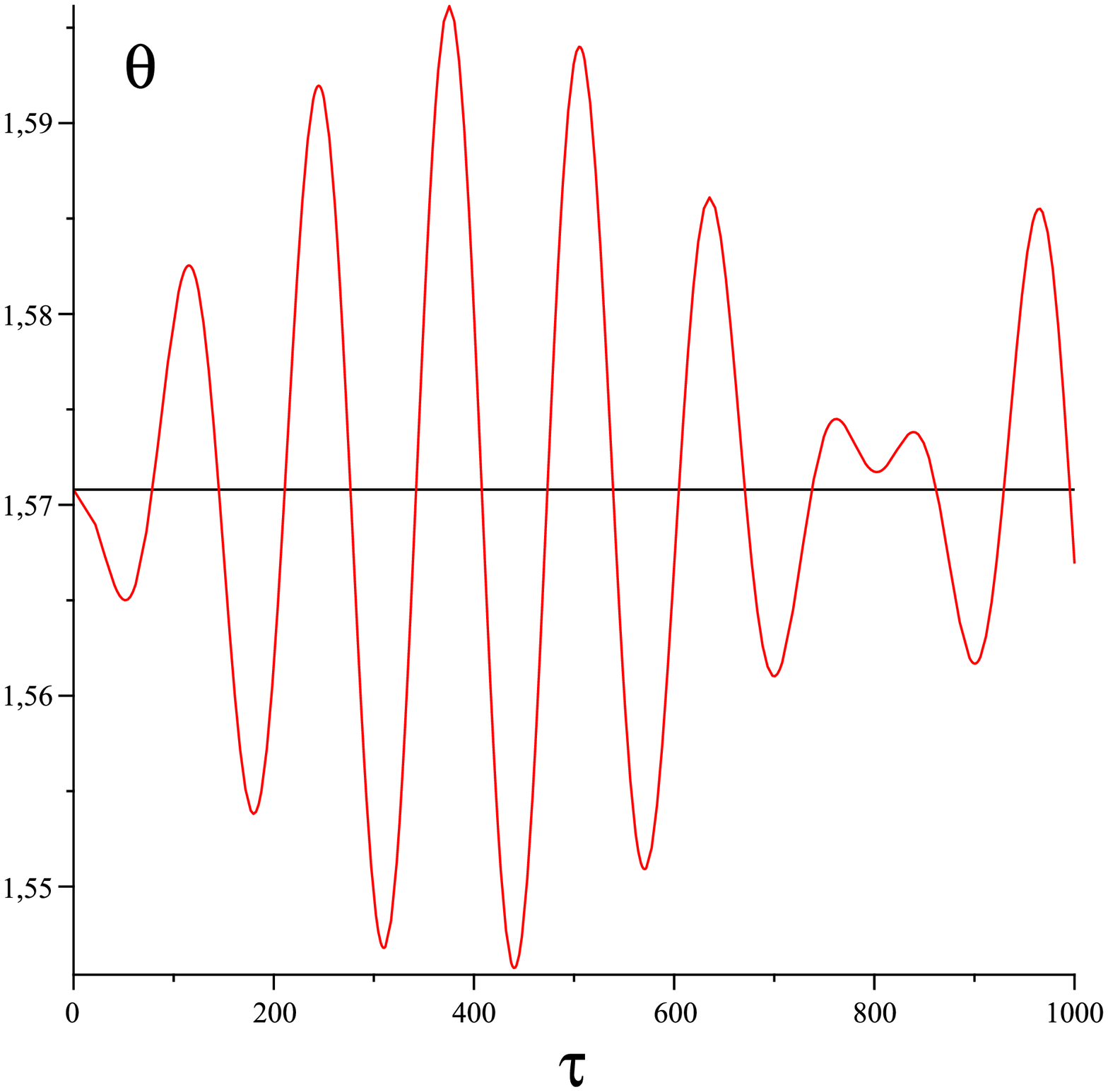}\\[.4cm]
\quad\mbox{(a)}\quad &\quad \mbox{(b)}\\[.6cm]
\includegraphics[scale=0.25]{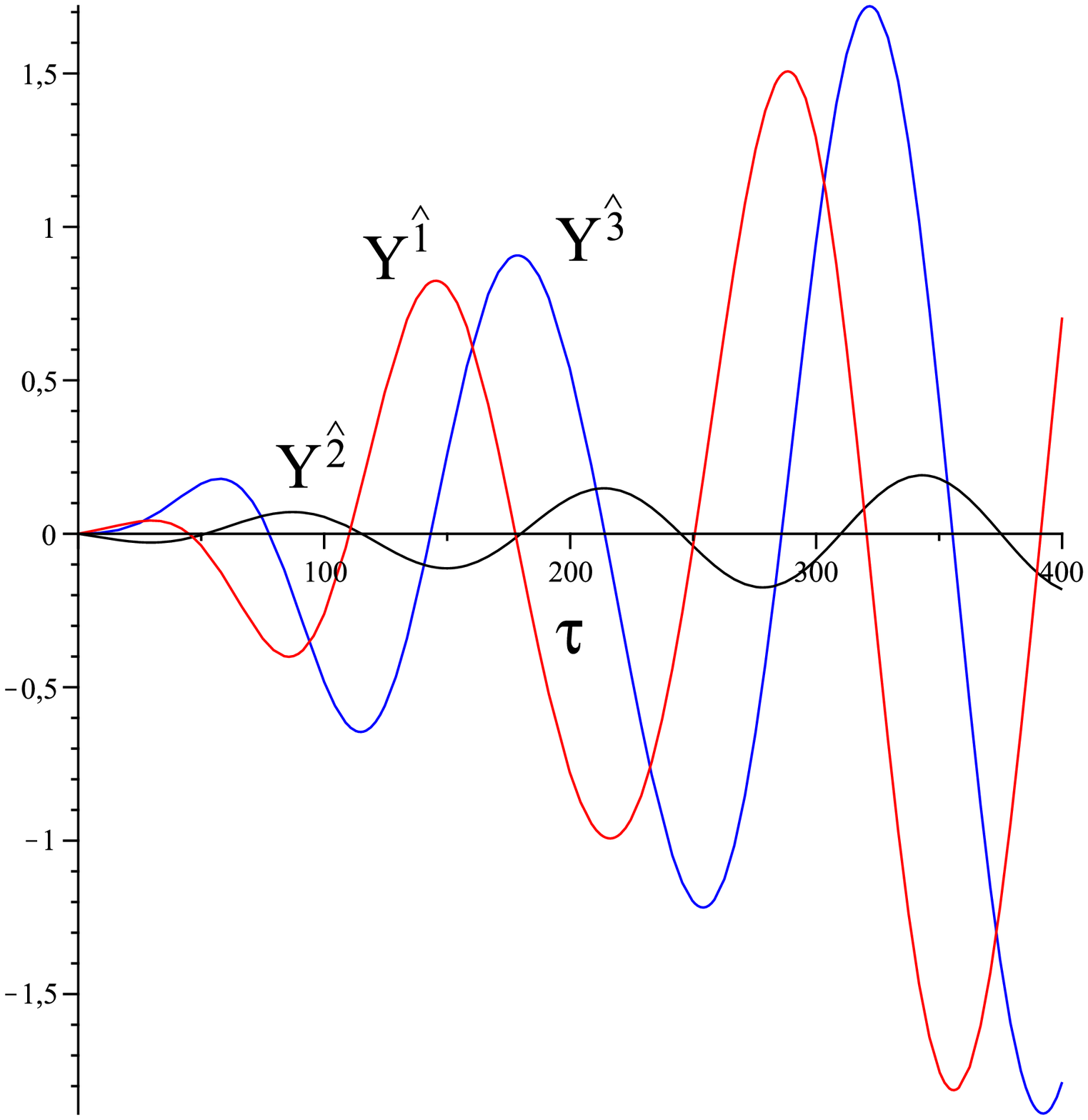}&\quad
\includegraphics[scale=0.25]{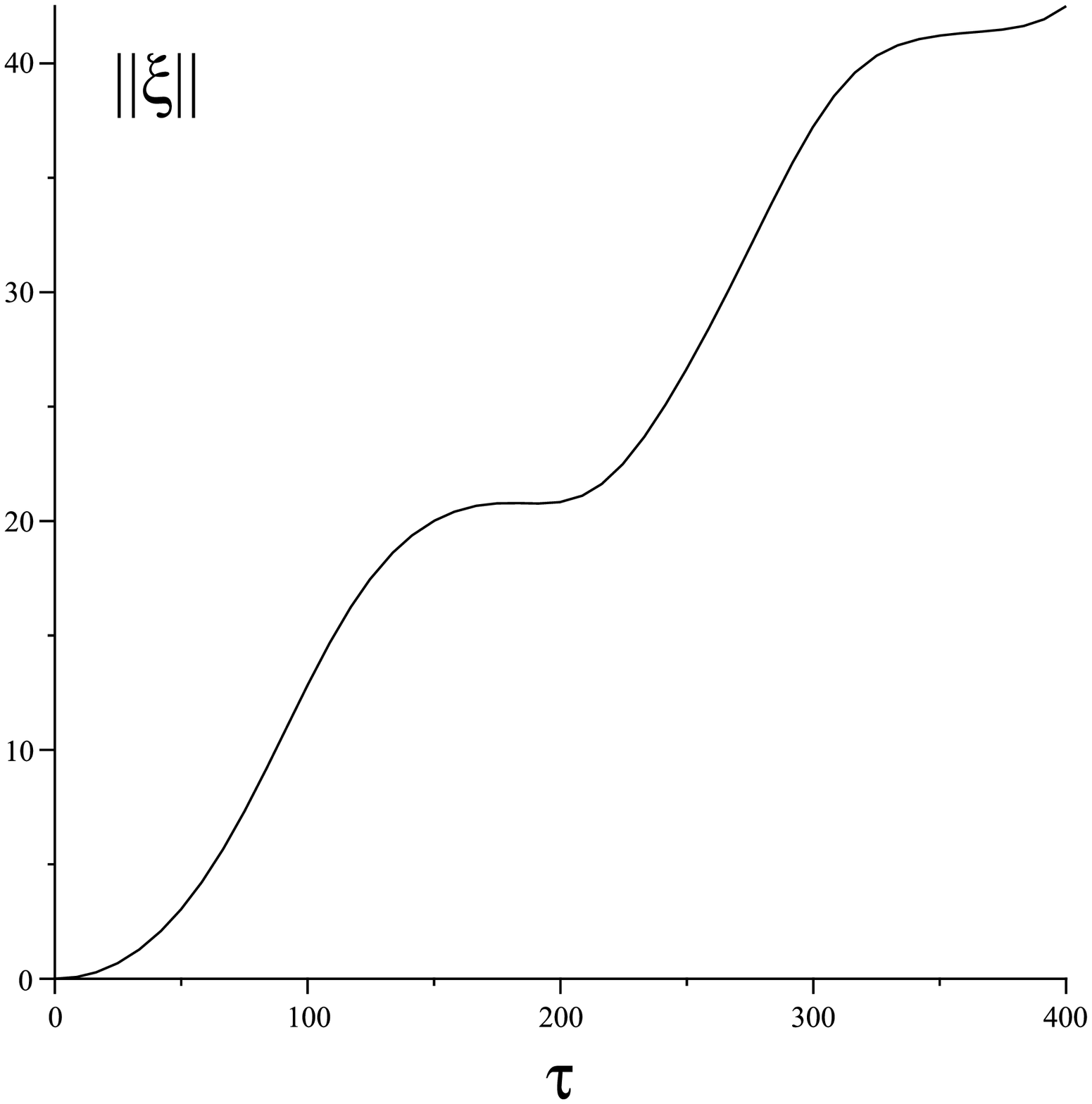}\\[.4cm]
\quad\mbox{(c)}\quad &\quad \mbox{(d)}
\end{array}$\\
\end{center}
\caption{Reference path: equatorial circular geodesic.\\
Panels (a) and (b) show the behavior of the perturbed radius and polar angle (red curves) as functions of the proper time, respectively.
Panel (c) shows the behavior of the frame components of the deviation vector $Y$ as functions of the proper time ($Y^{\hat 1}$ red, $Y^{\hat 2}$ black, $Y^{\hat 3}$ blue).
Panel (d) shows the behavior of the magnitude of the displacement vector $\xi$ as a function of the proper time.
The associated reference geodesic is an equatorial circular orbit at $r_0=8M$, so that $E\approx0.94$, $L\approx3.32$ and $K\approx8.1$, the black hole rotation parameter being set to $a/M=0.5$.
The spin parameter (only necessary for Figs. (a) and (b)) has been chosen as $\sigma/M=0.05$ and the spin orientation has the general orientation $N^{\hat 1}=0.5=N^{\hat 3}$, $N^{\hat 2}\approx-0.71$.
Note that the selected value for $\sigma$, even if representing a reference value, is not so far from actual astrophysical ones (see discussion at the end of Section II). 
The choice of initial conditions is the following: $t_{\rm(g)}(0)=0$, $r_{\rm(g)}(0)=8M$, $\theta_{\rm(g)}(0)=\pi/2$, $\phi_{\rm(g)}(0)=0$, $\xi^{\hat a}(0)=0=\rmd\xi^{\hat a}/\rmd\tau_{\rm(g)}(0)$, $\psi(0)=0$.
}
\label{fig:1}
\end{figure}

% figure 2

\begin{figure} 
\typeout{*** EPS figure 2}
\begin{center}
$\begin{array}{cc}
\includegraphics[scale=0.25]{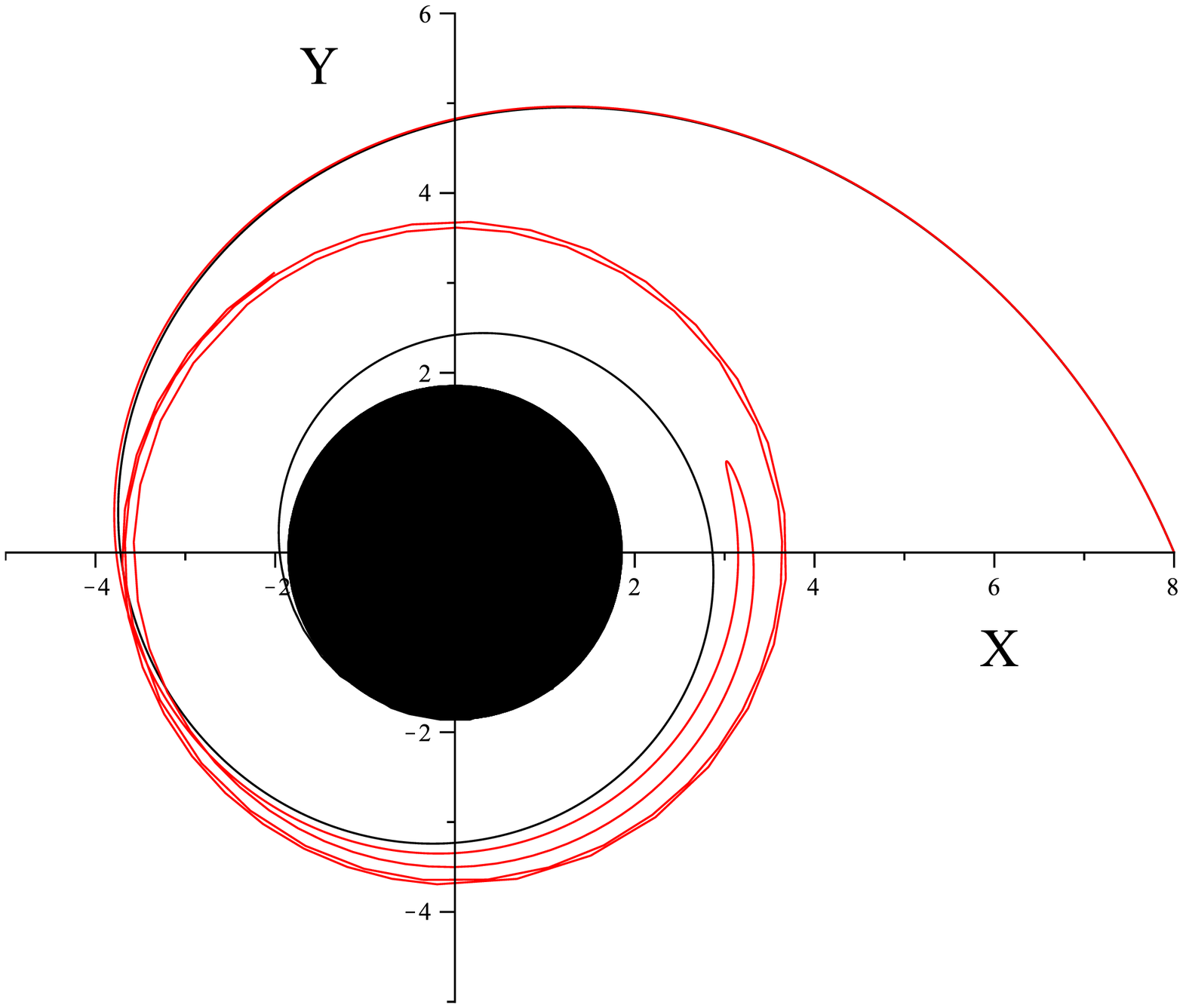}&\quad
\includegraphics[scale=0.23]{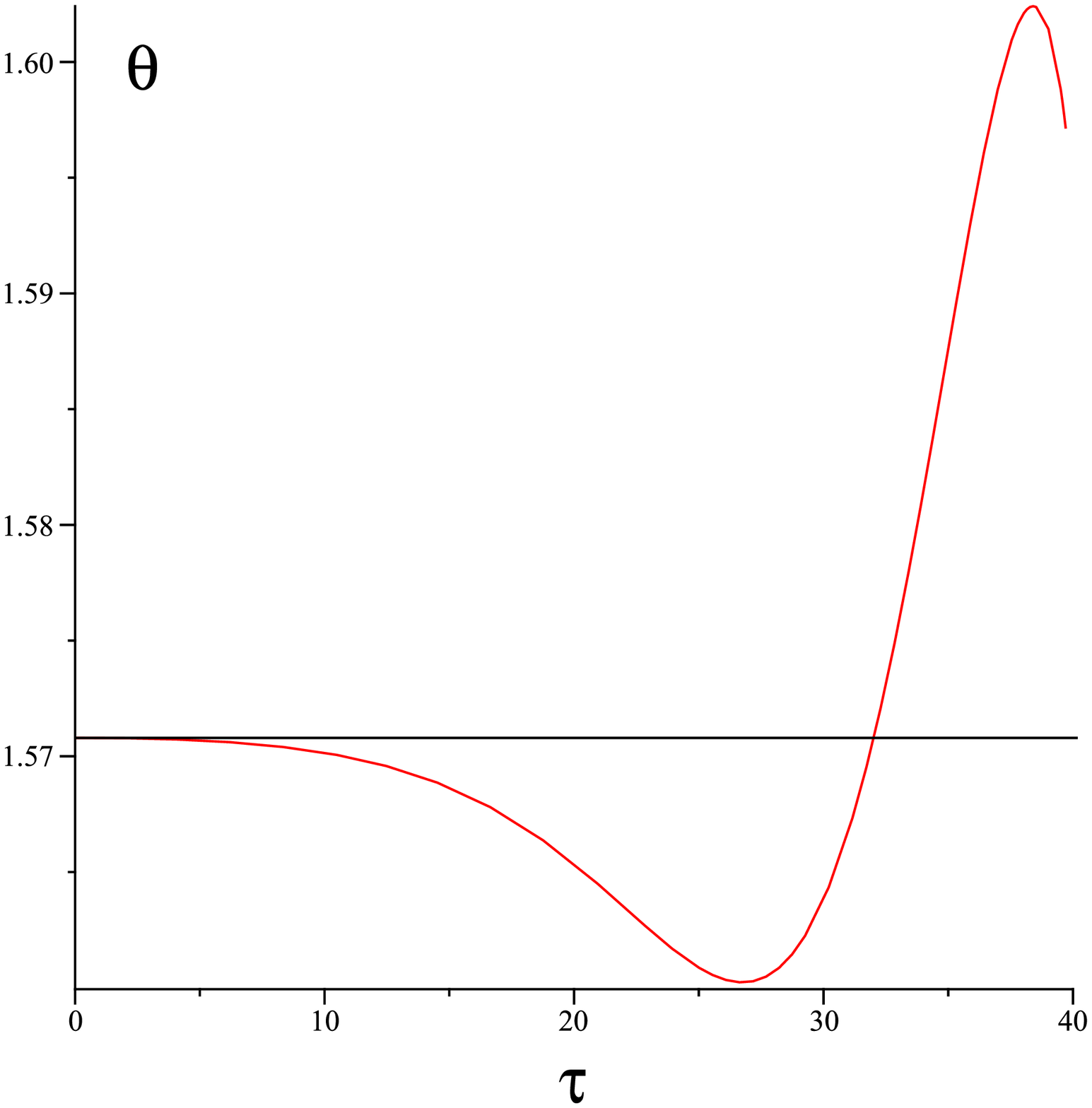}\\[.4cm]
\quad\mbox{(a)}\quad &\quad \mbox{(b)}\\[.6cm]
\includegraphics[scale=0.25]{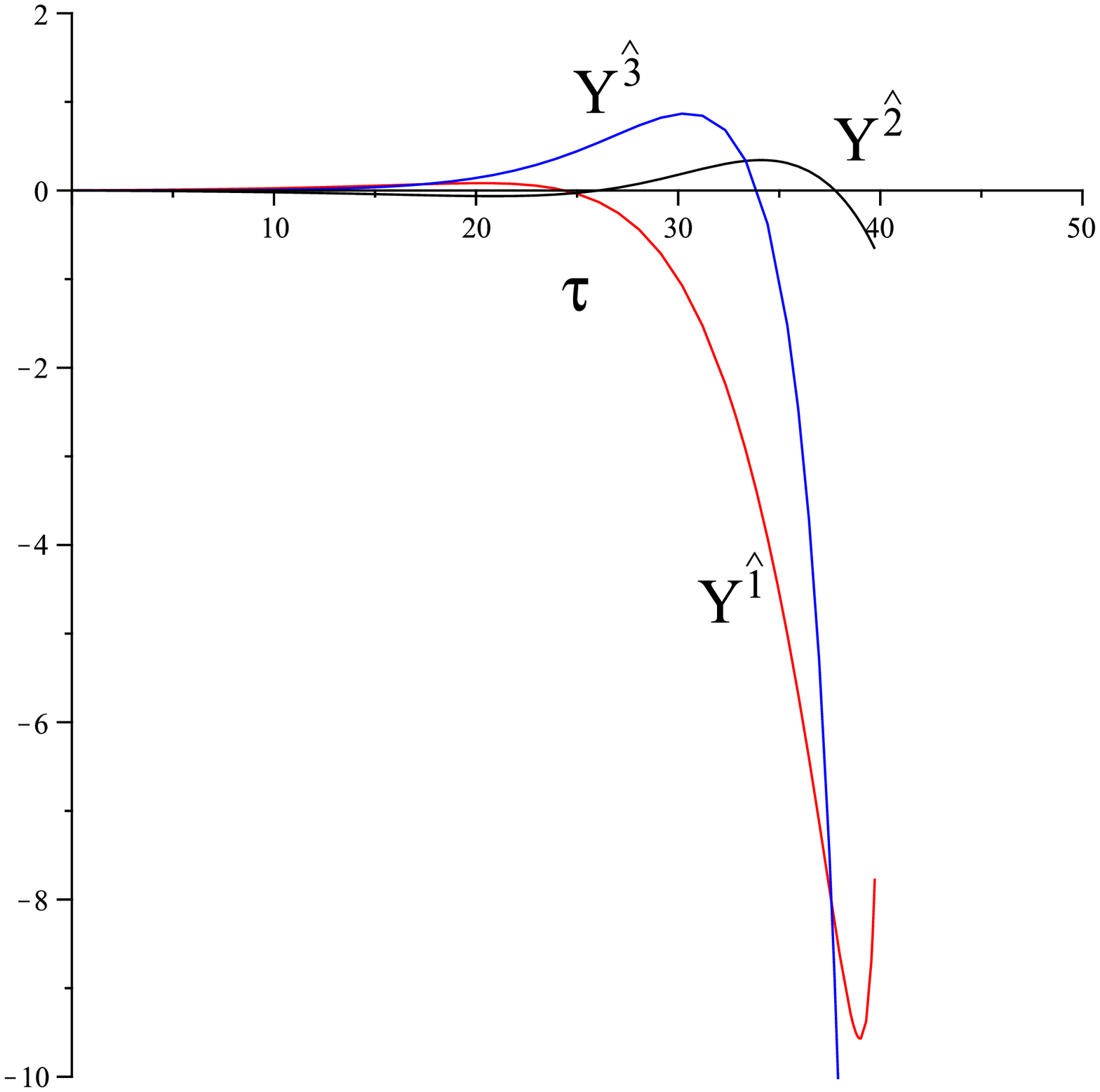}&\quad
\includegraphics[scale=0.25]{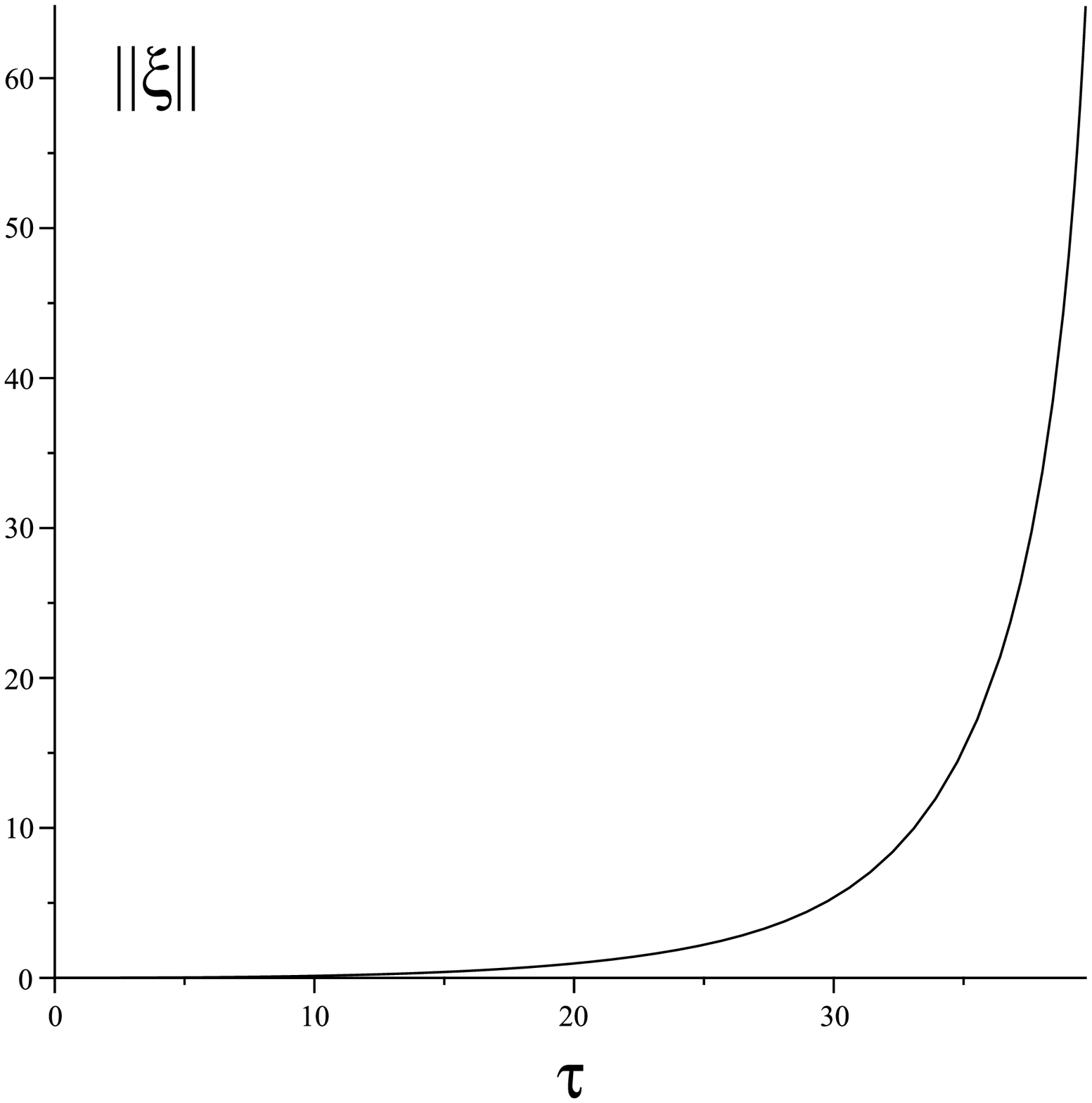}\\[.4cm]
\quad\mbox{(c)}\quad &\quad \mbox{(d)}
\end{array}$\\
\end{center}
\caption{Reference path: equatorial non-circular geodesic.\\
Panel (a) shows the projection of the perturbed trajectory on the $X-Y$ plane (red curve), where $X=r\cos\phi$ and $Y=r\sin\phi$ are Cartesian-like coordinates.
Panel (b) shows the behavior of the perturbed polar angle as a function of the proper time.
Panel (c) shows the behavior of the frame components of the deviation vector $Y$ as functions of the proper time ($Y^{\hat 1}$ red, $Y^{\hat 2}$ black, $Y^{\hat 3}$ blue).
Panel (d) shows the behavior of the magnitude of the displacement vector $\xi$ as a function of the proper time.
The parameters of the reference equatorial geodesic are chosen so that $E=0.95$, $L=3.1$, $K\approx6.9$ and $\epsilon_r=-1$.
The values of black hole rotation parameter, spin orientation and spin parameter as well as initial conditions are the same as in Fig. \ref{fig:1}.
The geodesic path crosses the horizon (black disk at $r_+\approx1.867M$) at $\tau_{\rm(g)}\approx 39.7$, whereas the path of the spinning particle becomes nearly circular at $r\approx 3.71M$. Correspondingly the deviation components take the values $Y^{\hat 1}\approx-7.56$, $Y^{\hat 2}\approx-0.67$ and $Y^{\hat 3}\approx-45.9$.
}
\label{fig:2}
\end{figure}

% figure 3

\begin{figure} 
\typeout{*** EPS figure 3}
\begin{center}
$\begin{array}{cc}
\includegraphics[scale=0.25]{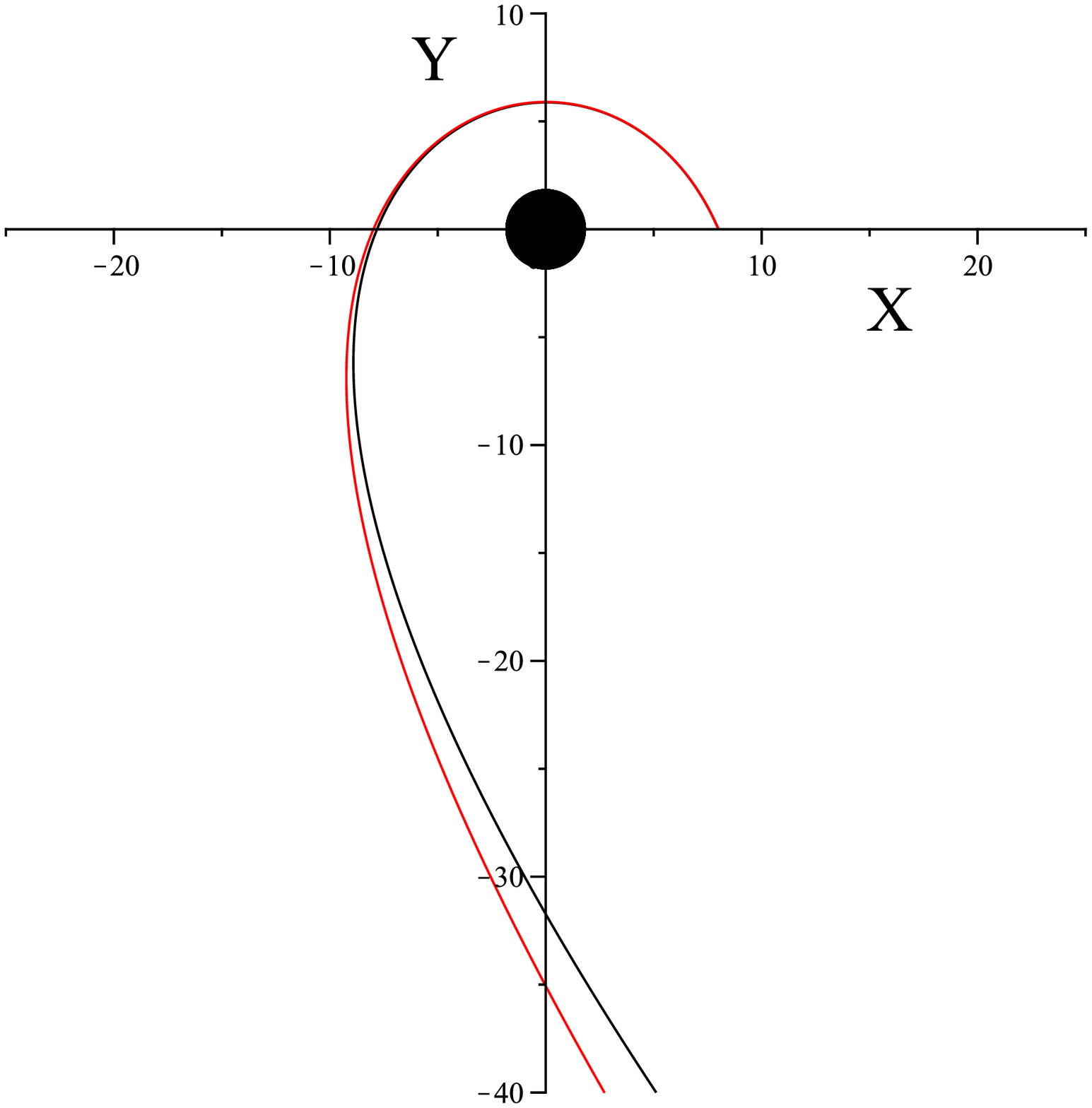}&\quad
\includegraphics[scale=0.25]{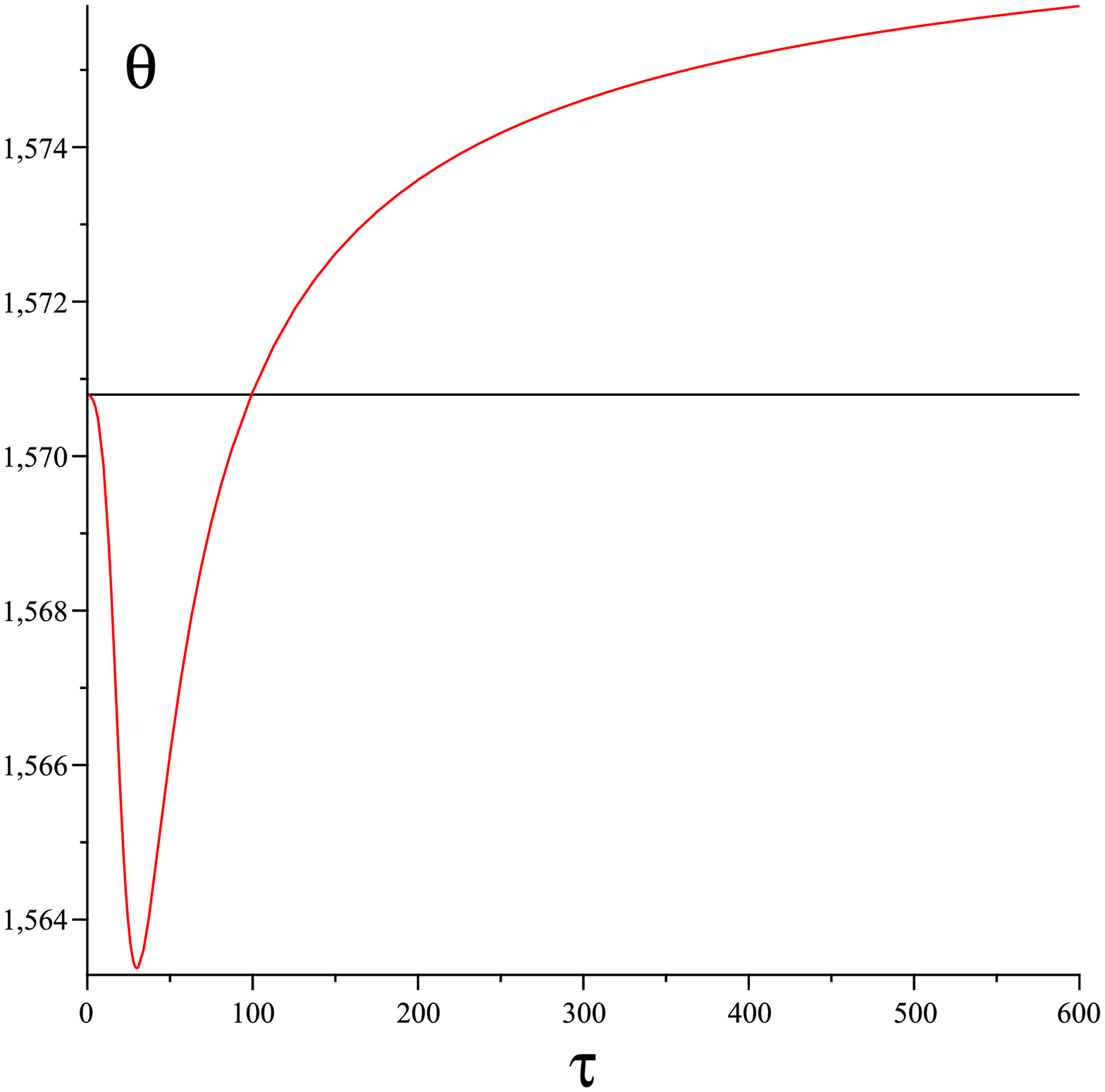}\\[.4cm]
\quad\mbox{(a)}\quad &\quad \mbox{(b)}\\[.6cm]
\includegraphics[scale=0.25]{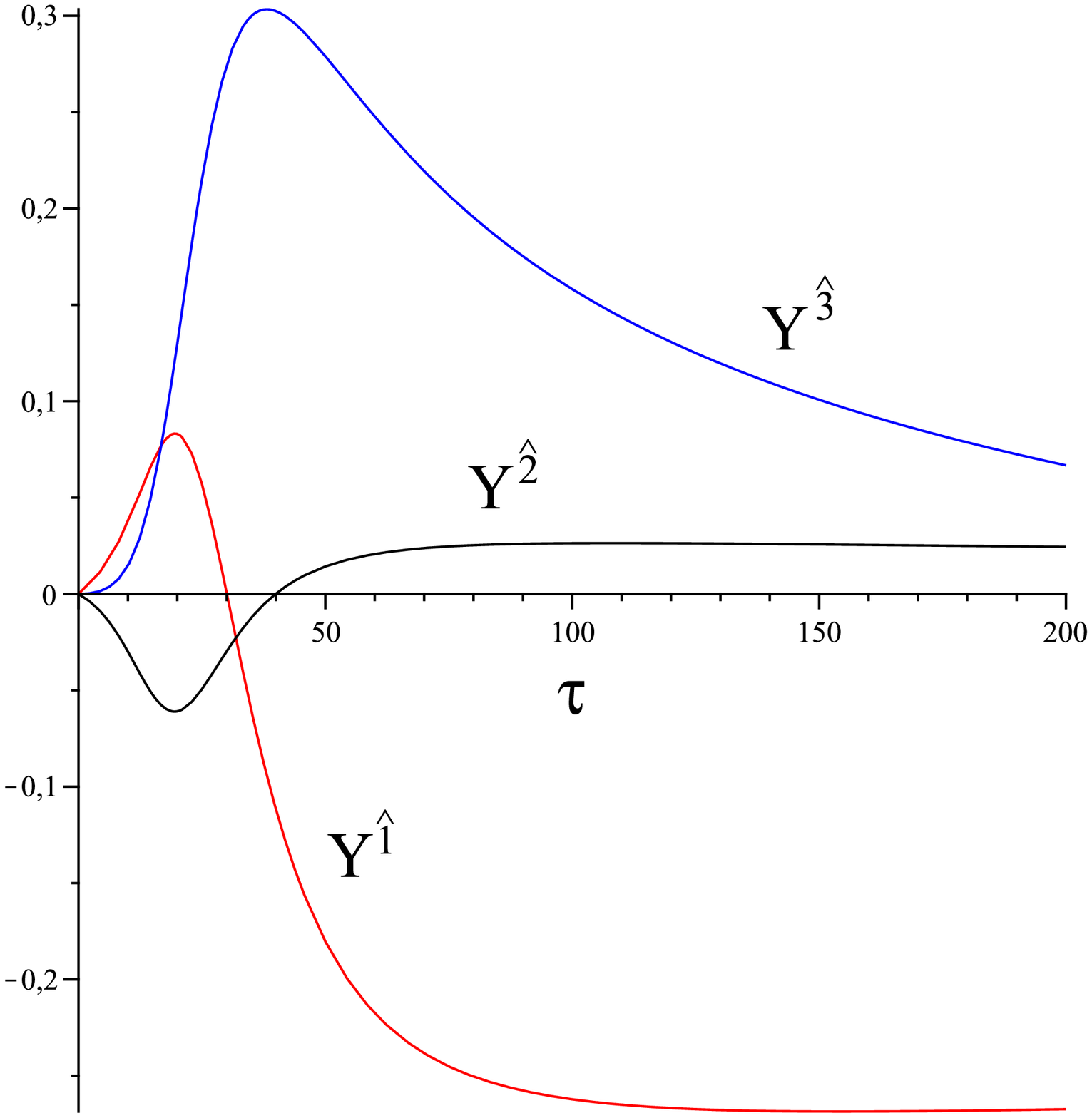}&\quad
\includegraphics[scale=0.25]{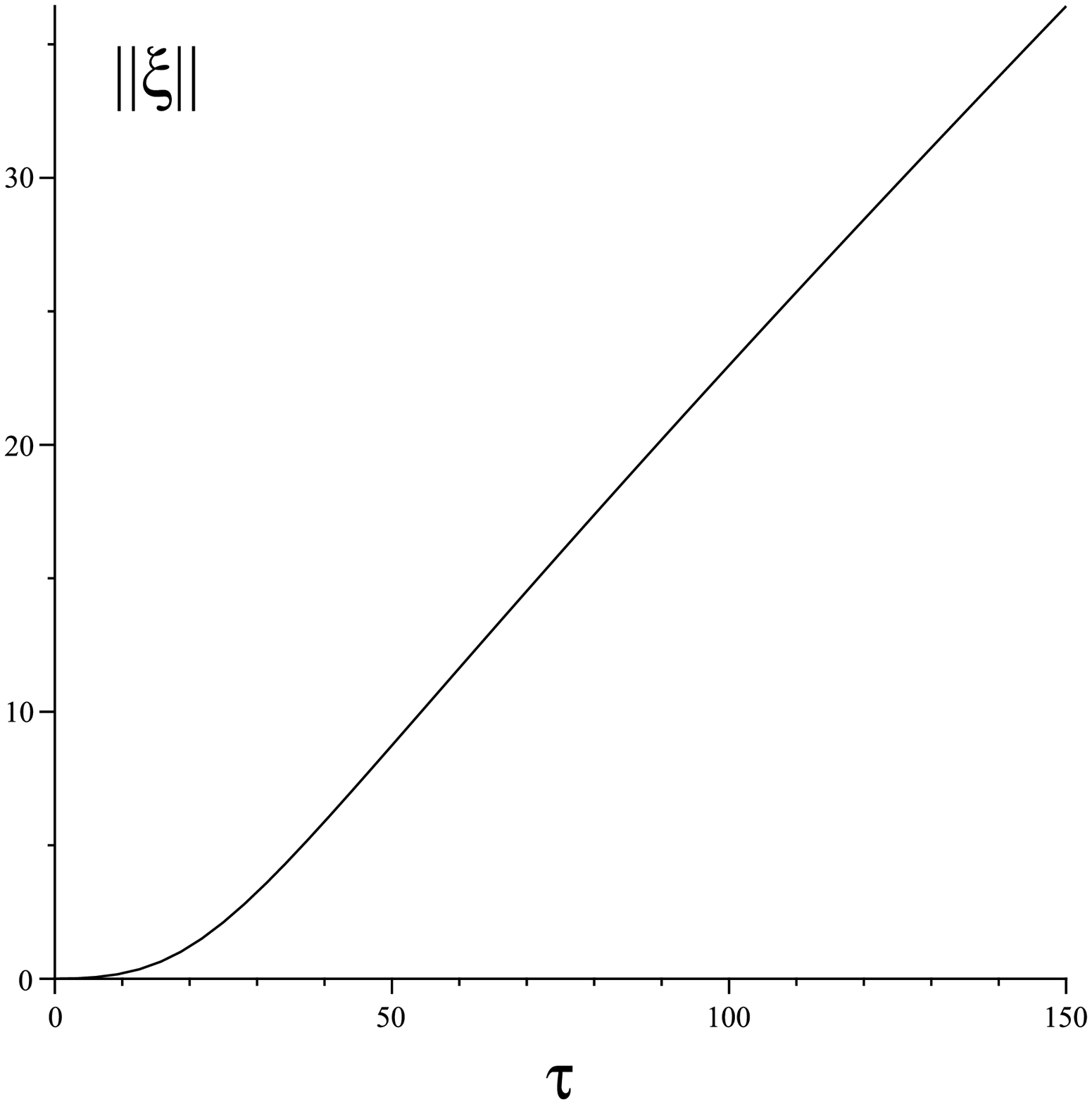}\\[.4cm]
\quad\mbox{(c)}\quad &\quad \mbox{(d)}
\end{array}$\\
\end{center}
\caption{Reference path: equatorial non-circular geodesic.\\
Panel (a) shows the projection of the perturbed trajectory on the $X-Y$ plane (red curve).
Panel (b) shows the behavior of the perturbed polar angle as a function of the proper time.
Panel (c) shows the behavior of the frame components of the deviation vector $Y$ as functions of the proper time ($Y^{\hat 1}$ red, $Y^{\hat 2}$ black, $Y^{\hat 3}$ blue).
Panel (d) shows the behavior of the magnitude of the displacement vector $\xi$ as a function of the proper time.
The parameters of the reference equatorial geodesic are chosen so that $E=0.95$, $L=5$, $K\approx12.4$ and $\epsilon_r=-1$.
The values of black hole rotation parameter, spin orientation and spin parameter as well as initial conditions are the same as in Fig. \ref{fig:1}.
}
\label{fig:3}
\end{figure}

% figure 4

\begin{figure} 
\typeout{*** EPS figure 4}
\begin{center}
$\begin{array}{cc}
\includegraphics[scale=0.25]{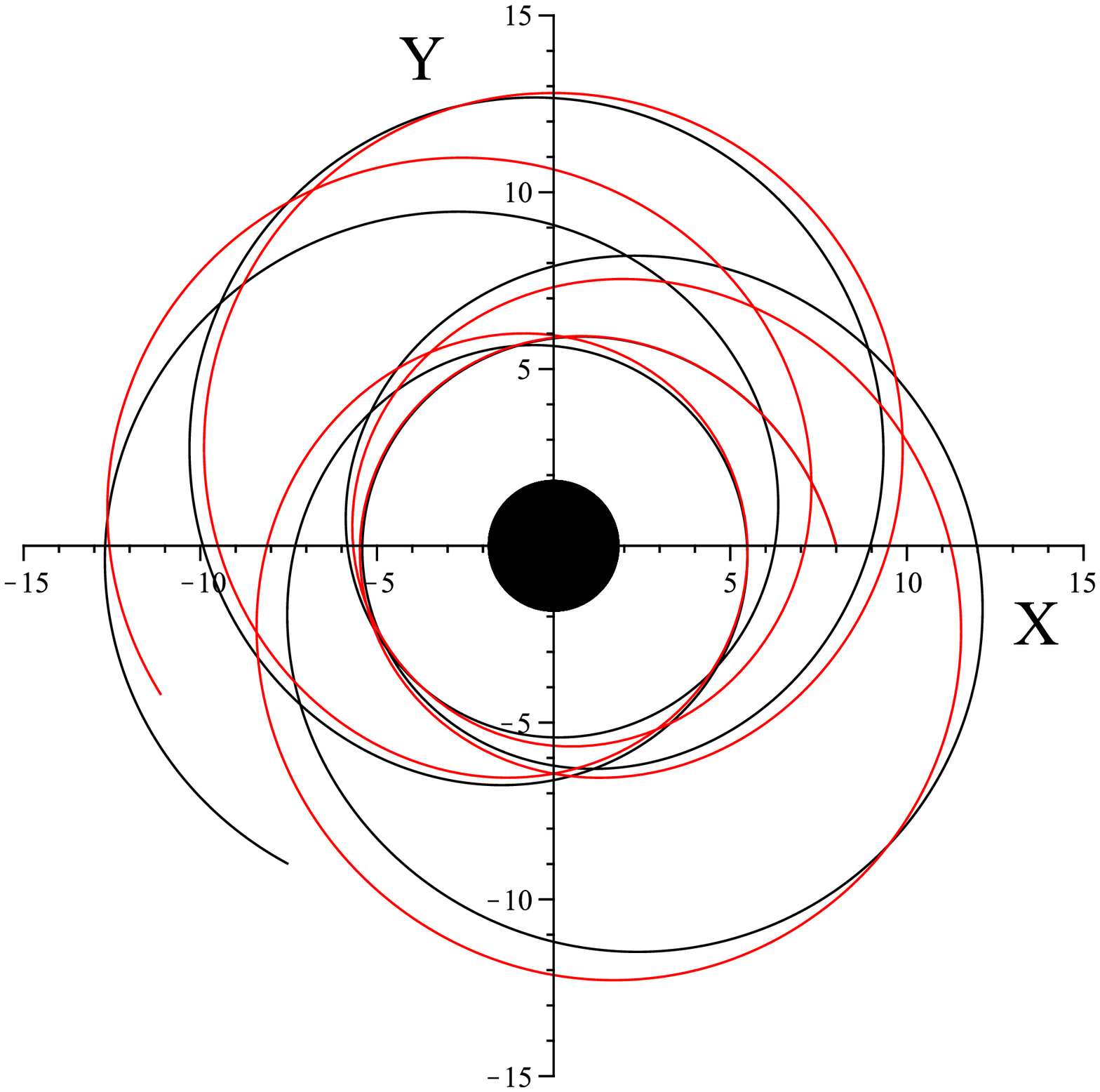}&\quad
\includegraphics[scale=0.25]{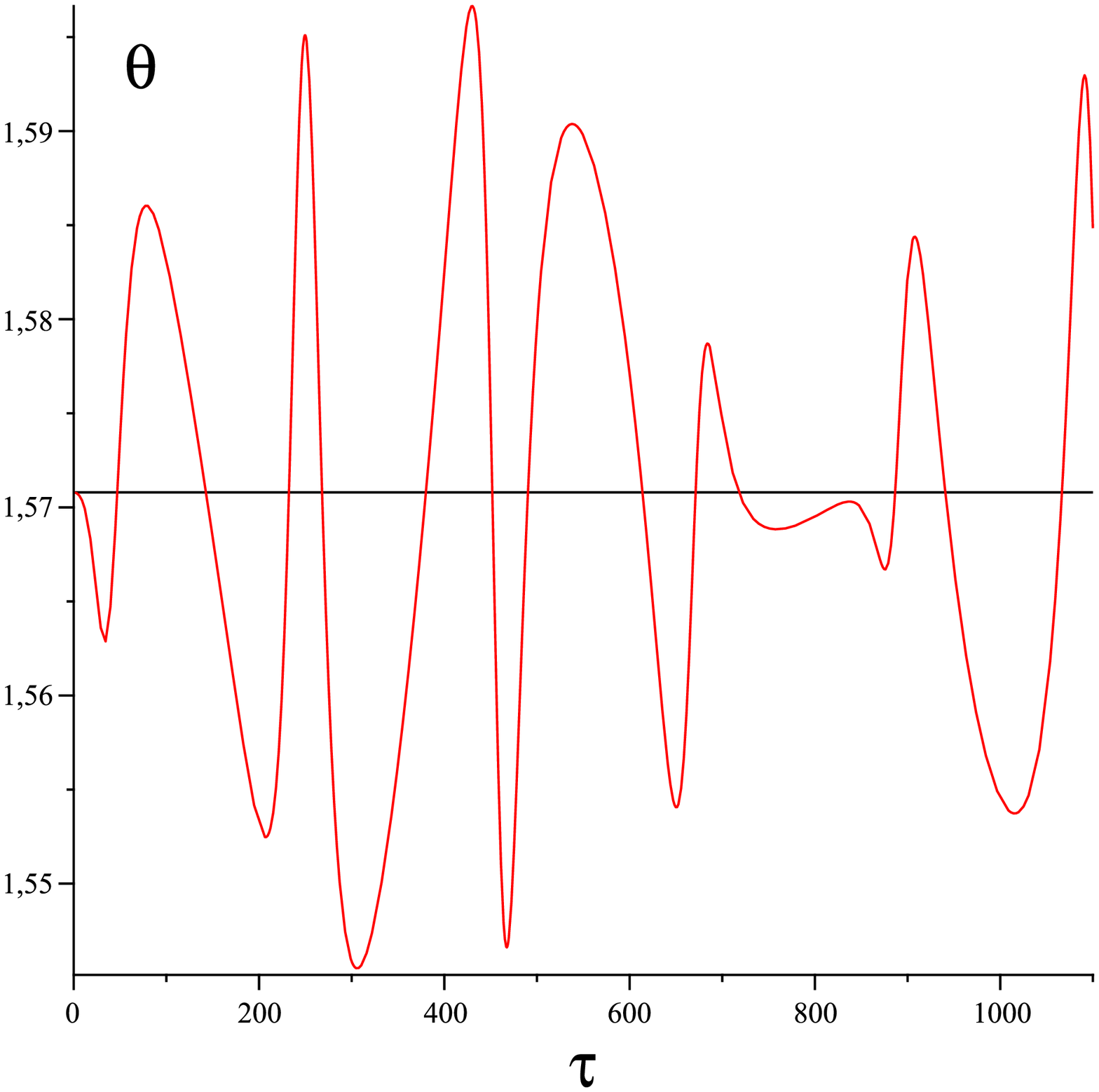}\\[.4cm]
\quad\mbox{(a)}\quad &\quad \mbox{(b)}\\[.6cm]
\includegraphics[scale=0.25]{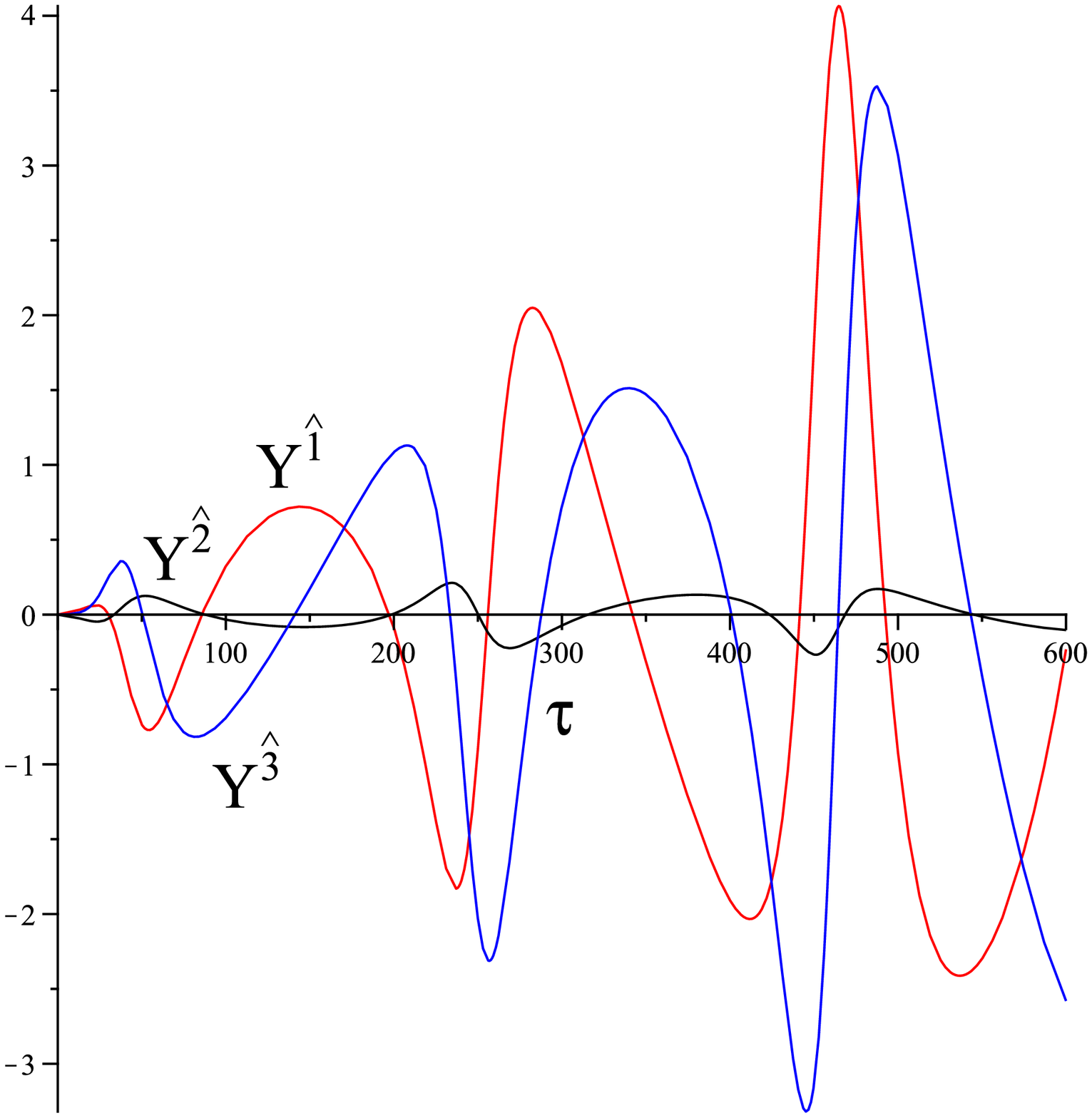}&\quad
\includegraphics[scale=0.25]{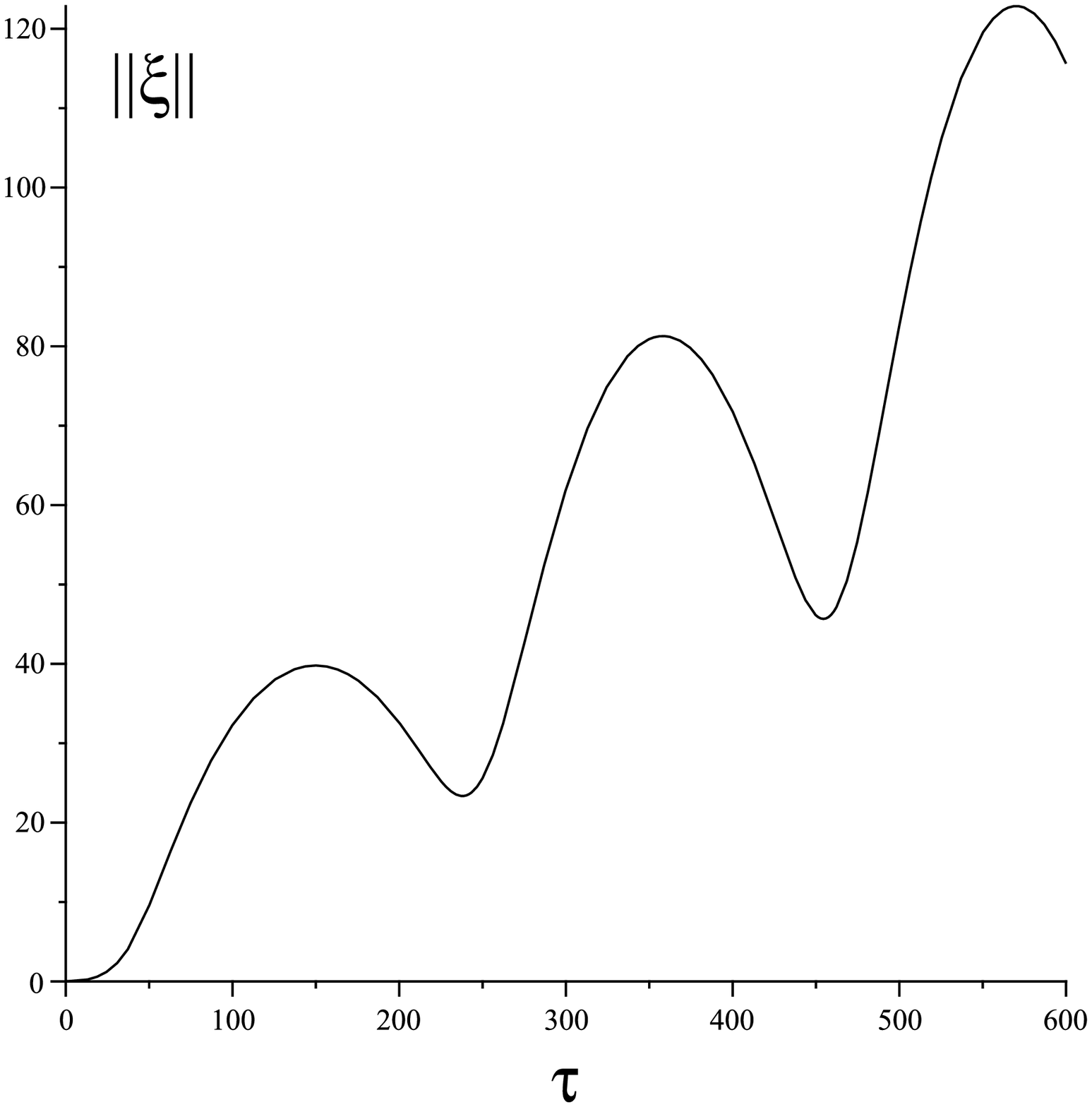}\\[.4cm]
\quad\mbox{(c)}\quad &\quad \mbox{(d)}
\end{array}$\\
\end{center}
\caption{Reference path: equatorial non-circular geodesic.\\
Panels (a) shows the projection of the perturbed trajectory on the $X-Y$ plane (red curve).
Panels (b) shows the behavior of the perturbed polar angle as a function of the proper time.
Panel (c) shows the behavior of the frame components of the deviation vector $Y$ as functions of the proper time ($Y^{\hat 1}$ red, $Y^{\hat 2}$ black, $Y^{\hat 3}$ blue).
Panel (d) shows the behavior of the magnitude of the displacement vector $\xi$ as a function of the proper time.
The parameters of the reference equatorial geodesic are chosen so that $E=0.95$, $L=3.3$, $K\approx8$ and $\epsilon_r=-1$.
The values of black hole rotation parameter, spin orientation and spin parameter as well as initial conditions are the same as in Fig. \ref{fig:1}.
}
\label{fig:4}
\end{figure}

% figure 5

\begin{figure} 
\typeout{*** EPS figure 5}
\begin{center}
$\begin{array}{cc}
\includegraphics[scale=0.25]{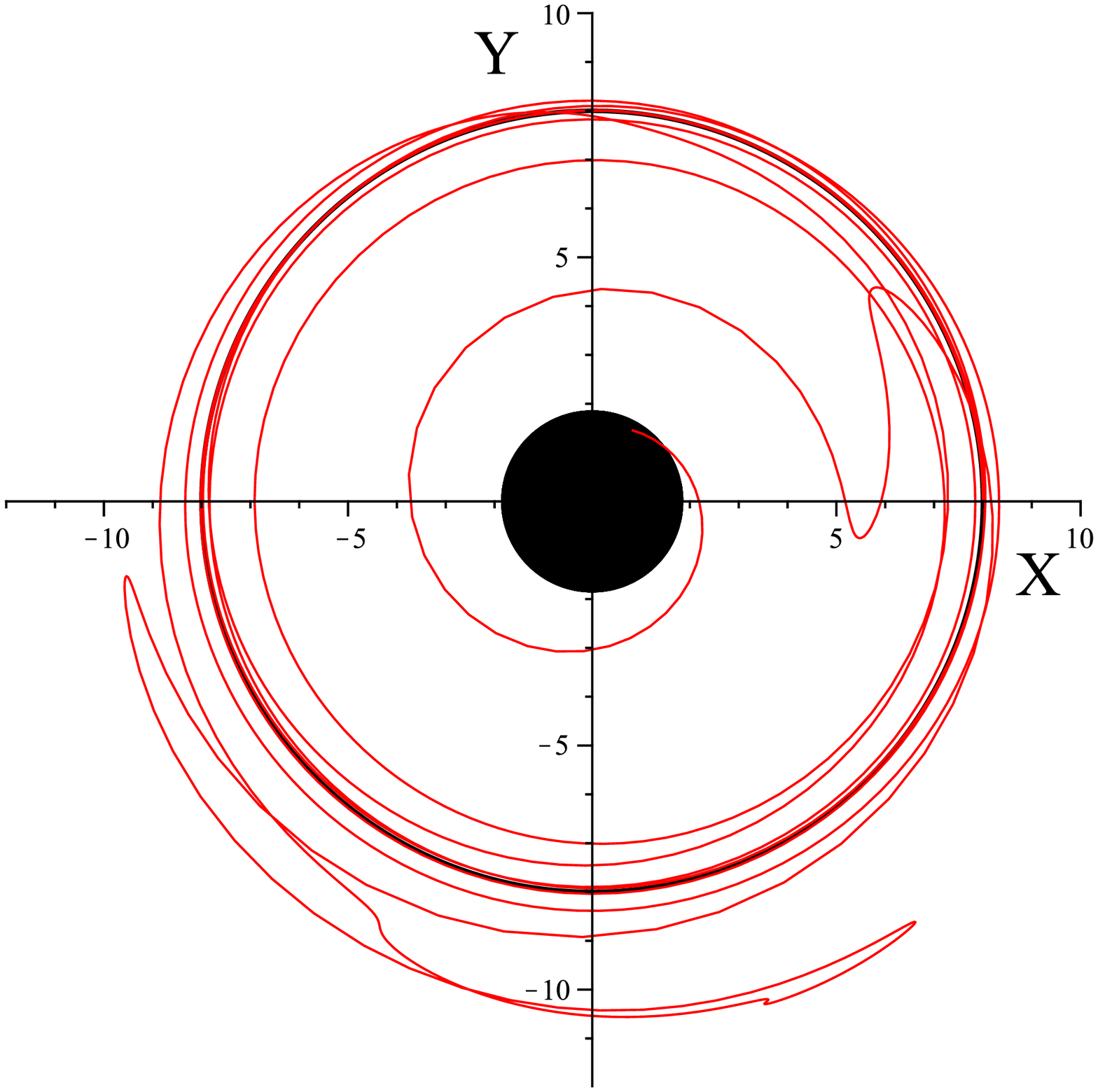}&\quad
\includegraphics[scale=0.25]{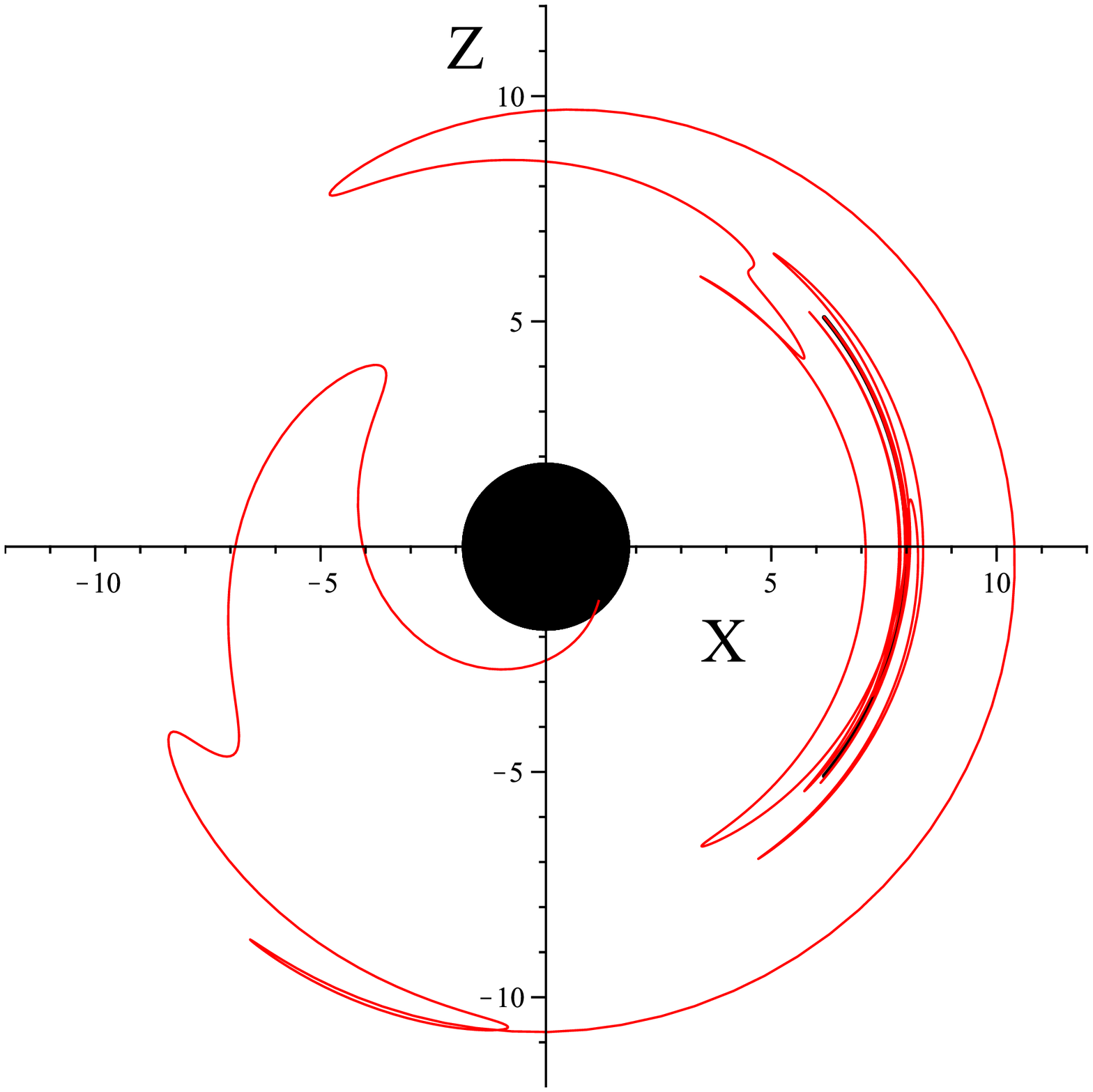}\\[.4cm]
\quad\mbox{(a)}\quad &\quad \mbox{(b)}\\[.6cm]
\includegraphics[scale=0.25]{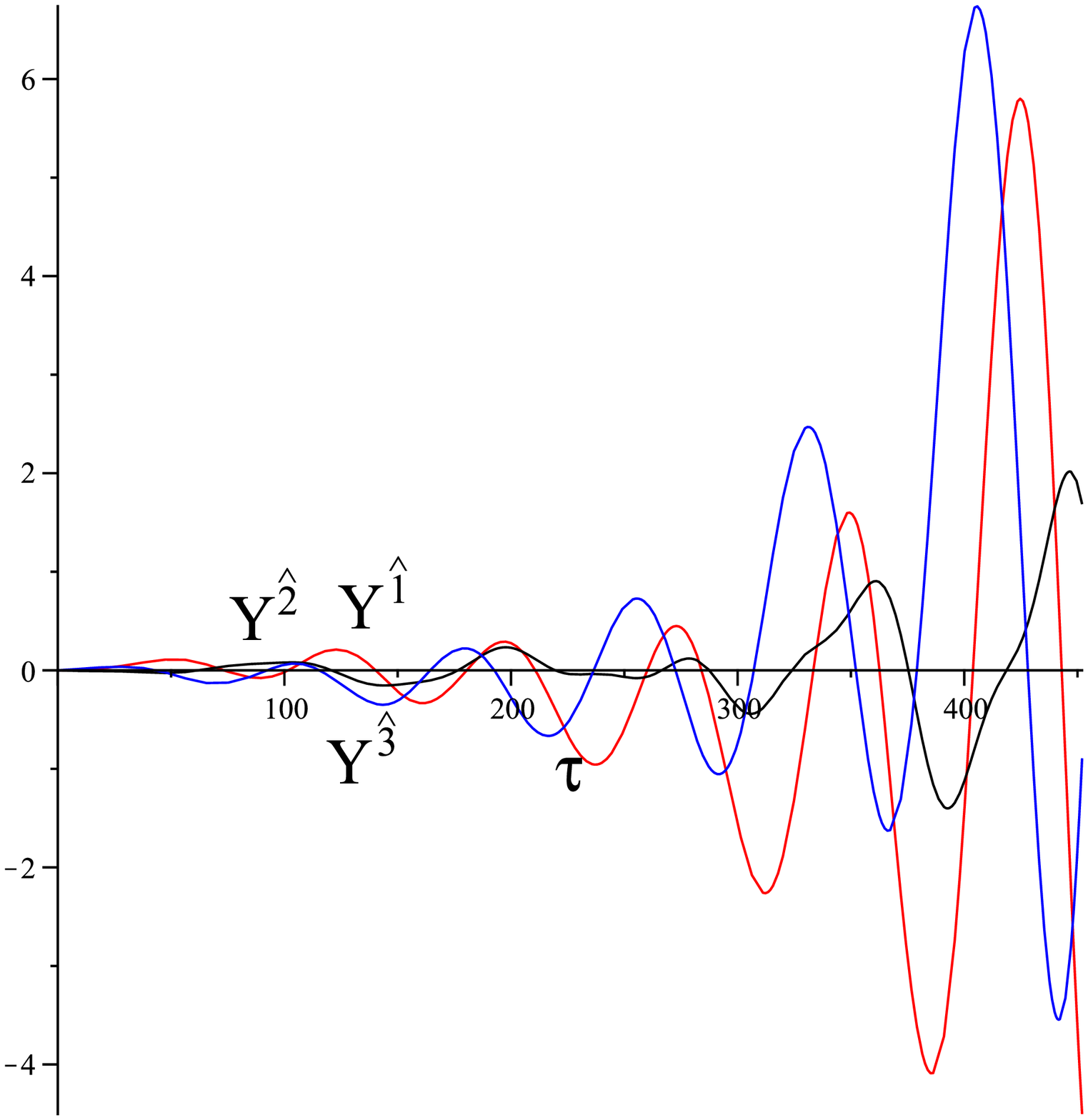}&\quad
\includegraphics[scale=0.25]{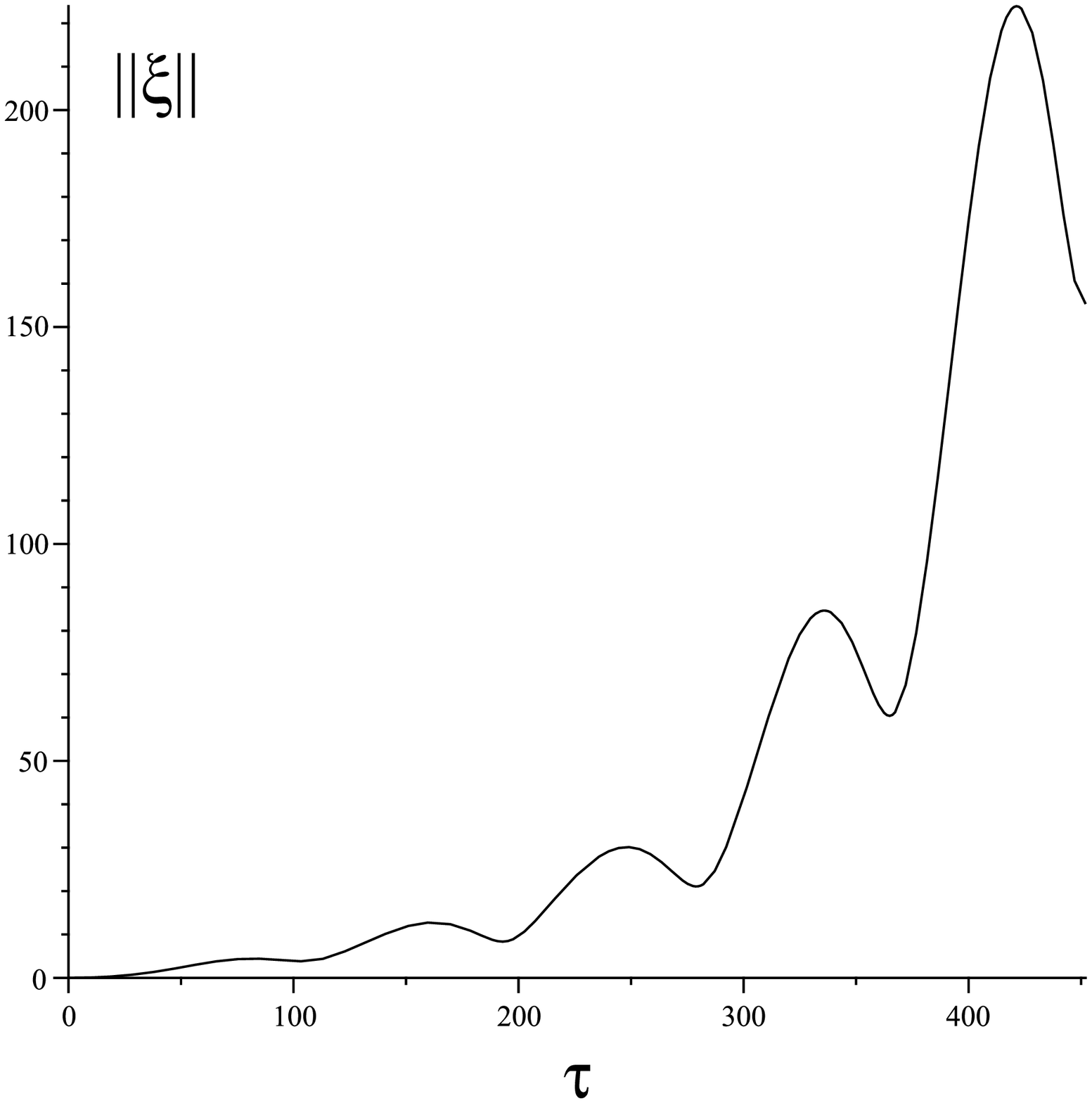}\\[.4cm]
\quad\mbox{(c)}\quad &\quad \mbox{(d)}
\end{array}$\\
\end{center}
\caption{Reference path: non-equatorial circular geodesic.\\
Panels (a) and (b) show the projections of the perturbed trajectory on the $X-Y$ plane and $X-Z$ plane (red curves), respectively, where $X=r\sin\theta\cos\phi$, $Y=r\sin\theta\sin\phi$ and $Z=r\cos\theta$ are Cartesian-like coordinates.
Panel (c) shows the behavior of the frame components of the deviation vector $Y$ as functions of the proper time ($Y^{\hat 1}$ red, $Y^{\hat 2}$ black, $Y^{\hat 3}$ blue).
Panel (d) shows the behavior of the magnitude of the displacement vector $\xi$ as a function of the proper time.
The background spin parameter has been set to $a/M=0.5$.
The associated reference geodesic is a spherical geodesic at $r_0=8M$ with $\epsilon_\theta=-1$ and $K=5$, so that $E\approx0.94$ and $L\approx5.56$.
The values of black hole rotation parameter, spin orientation and spin parameter as well as initial conditions are the same as in Fig. \ref{fig:1}.
}
\label{fig:5}
\end{figure}

% figure 6

\begin{figure} 
\typeout{*** EPS figure 6}
\begin{center}
$\begin{array}{cc}
\includegraphics[scale=0.25]{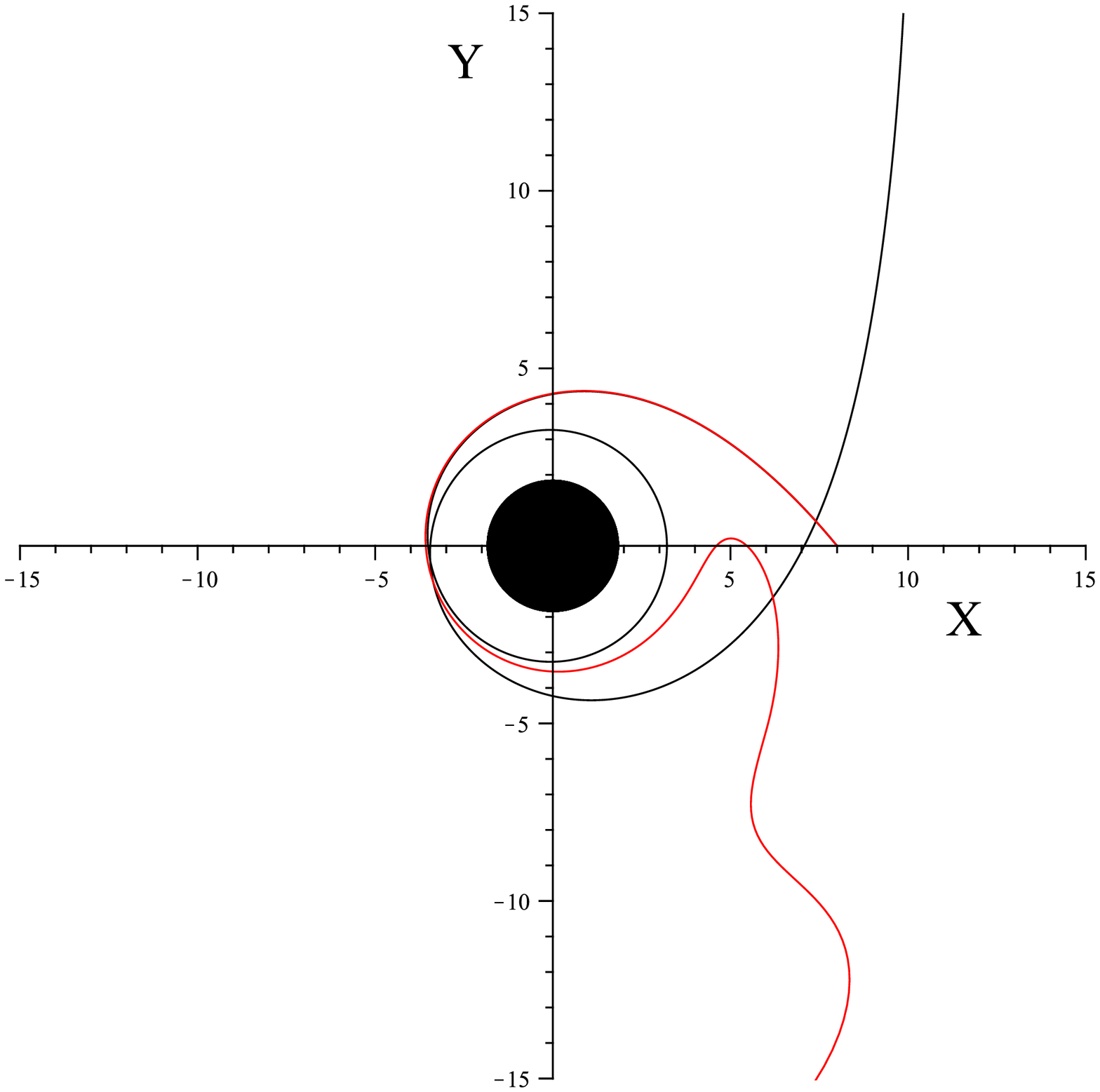}&\quad
\includegraphics[scale=0.25]{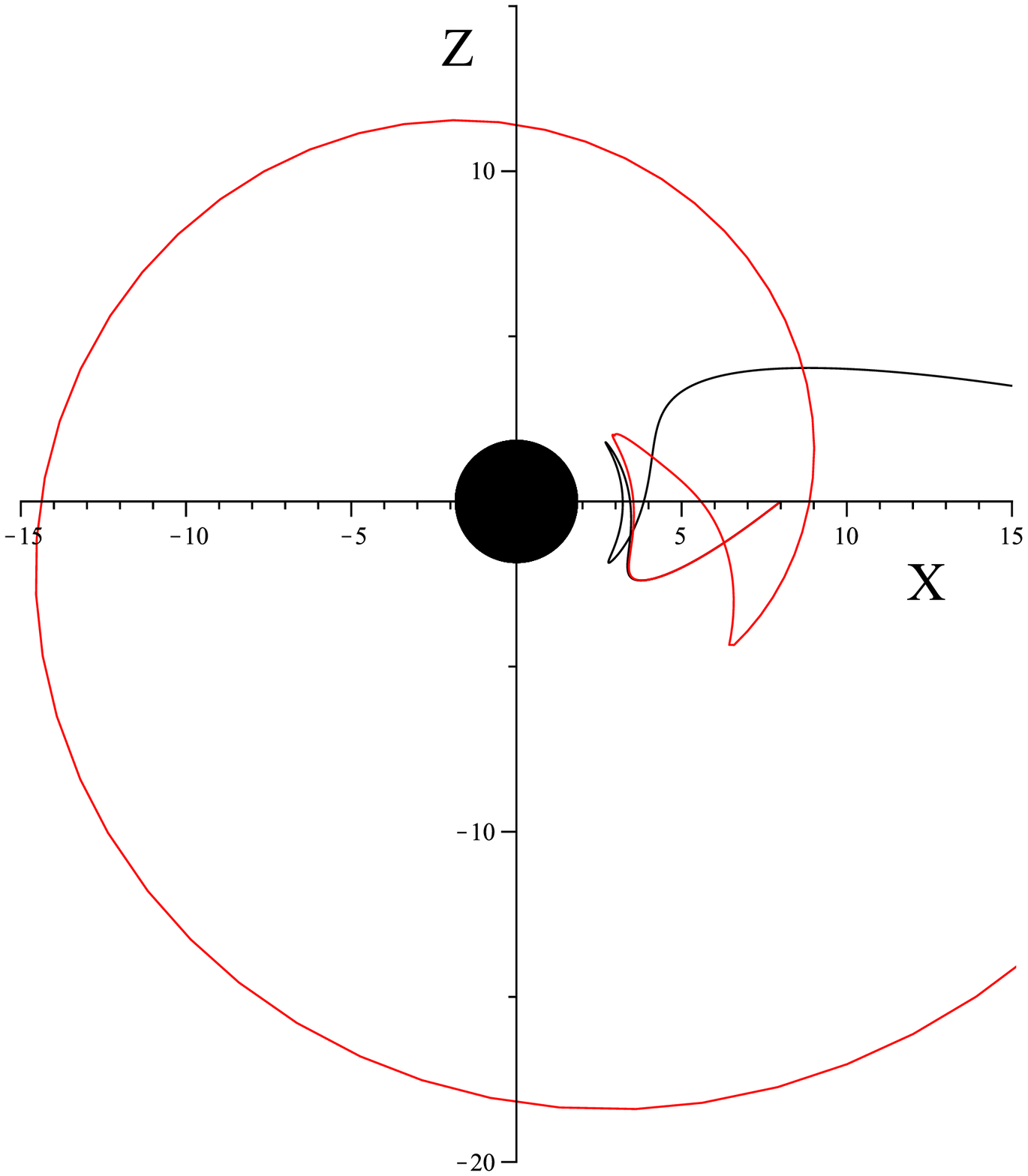}\\[.4cm]
\quad\mbox{(a)}\quad &\quad \mbox{(b)}\\[.6cm]
\includegraphics[scale=0.25]{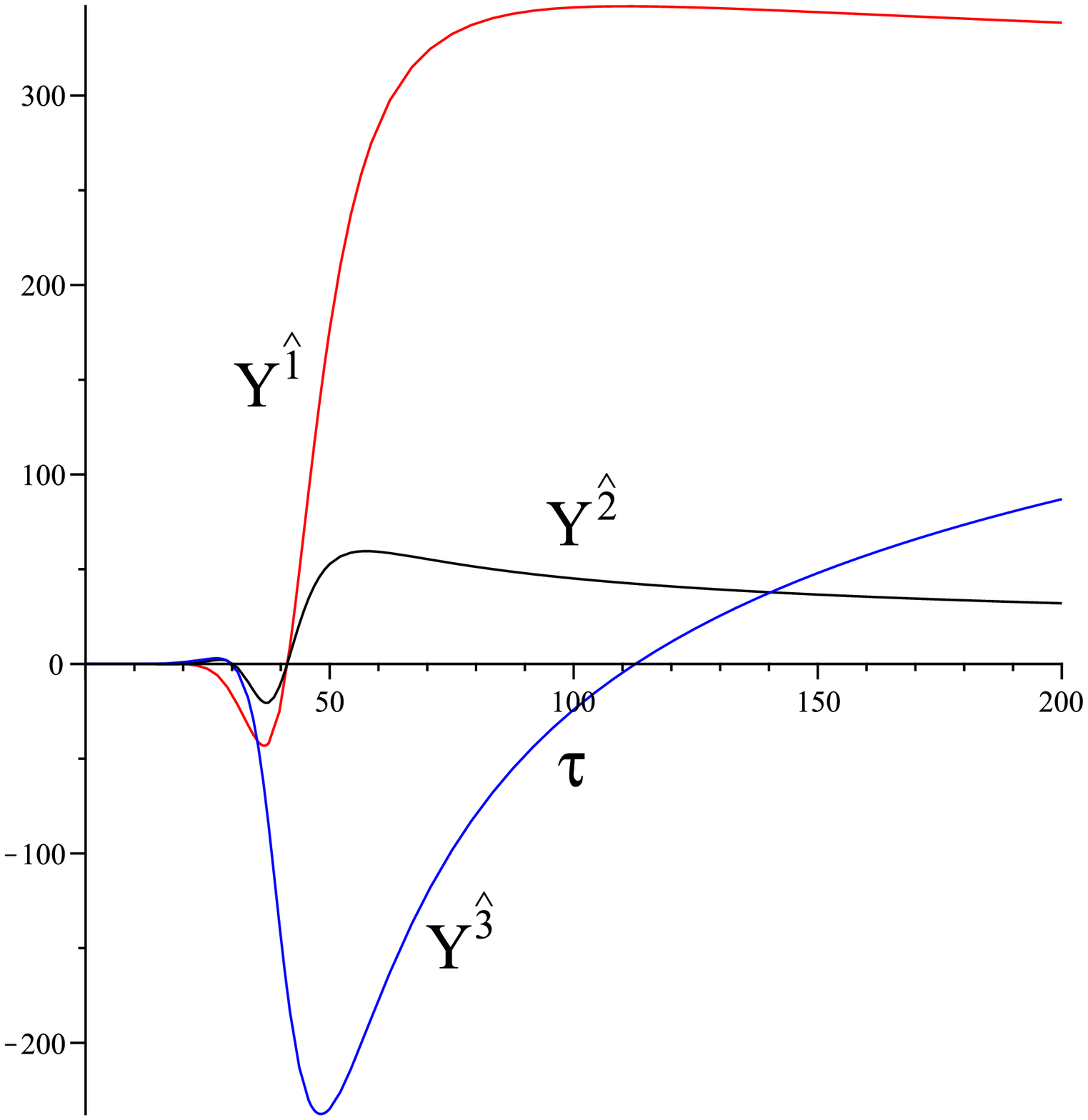}&\quad
\includegraphics[scale=0.25]{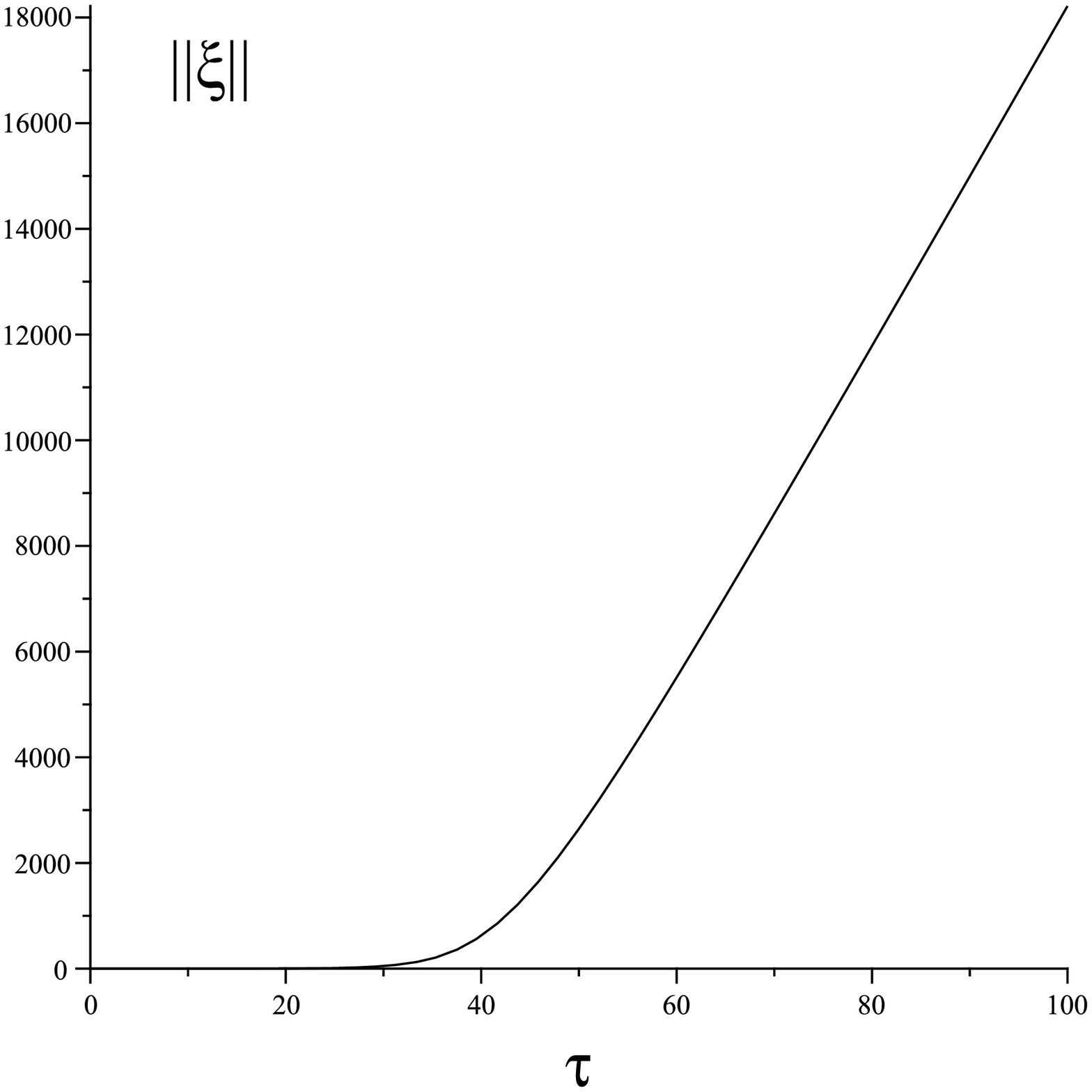}\\[.4cm]
\quad\mbox{(c)}\quad &\quad \mbox{(d)}
\end{array}$\\
\end{center}
\caption{
Reference path: non-equatorial non-circular geodesic.\\
Panels (a) and (b) show the projections of the perturbed trajectory on the $X-Y$ plane and $X-Z$ plane (red curves), respectively.
Panel (c) shows the behavior of the frame components of the deviation vector $Y$ as functions of the proper time ($Y^{\hat 1}$ red, $Y^{\hat 2}$ black, $Y^{\hat 3}$ blue).
Panel (d) shows the behavior of the magnitude of the displacement vector $\xi$ as a function of the proper time.
The parameters of the reference geodesic are chosen so that $E=0.9$, $L=3$, $K=10.5$, $\epsilon_r=-1$ and $\epsilon_\theta=1$.
The values of black hole rotation parameter, spin orientation and spin parameter as well as initial conditions are the same as in Fig. \ref{fig:1}.
}
\label{fig:6}
\end{figure}

\subsection{Discussion}

The study of spin-geodesic deviations addressed in the present section can be summarized as follows.
When considering an equatorial circular geodesic as the reference path (see Fig. \ref{fig:1}) the deviations due to spin remain small for a large interval of proper time, and exhibit an oscillating behavior with specific frequencies analytically determined in Ref. \cite{mashsingh}. As expected from the presence of a secular drift term in the solution for the perturbed azimuthal motion, the magnitude of the oscillations increases with time, but very slowly, as shown in panel (d), where the behavior of the magnitude of the displacement vector is plotted as a function of the proper time.
As a consequence, the perturbed orbit remains quasi-circular and quasi-equatorial even after many revolutions.  

If the reference orbit is still an equatorial geodesic, but non-circular as in Figs. \ref{fig:2}--\ref{fig:4}, the situation can be quite different.
For instance, in the case of the spiraling orbit of Fig. \ref{fig:2} the deviations are more enhanced approaching the horizon due to the increasing of the spin-curvature force there.
The initial oscillatory behavior of the deviation vector components is lost soon (see panel (c)), and just after a few revolutions the magnitude of the displacement vector increases of about 2 orders of magnitude, so that the perturbed orbit is very different from the reference one, as shown in panel (a).
In the case of the escape orbit of Fig. \ref{fig:3} we still have increasing deviations when the path is close to the hole, but these become less and less enhanced asymptotically far from the hole where the spin-curvature force vanishes.
As a consequence, the perturbed orbit mostly remains close to the geodesic path.
In Fig. \ref{fig:4} we have considered an equatorial non-circular geodesic with a \lq\lq rosetta-like" path.
As in the circular case we see oscillations in the variables $r$ and $\theta$, but now with a not well defined periodicity.
The perturbed orbit still remains close enough to the geodesic.

The case of non-equatorial reference geodesics has been investigated in Fig. \ref{fig:5} and \ref{fig:6}. 
In Fig. \ref{fig:5} we have selected a spherical geodesic. The components of the deviation vector are initially highly suppressed, but after a few revolutions they start increasing rapidly leading to a spiraling motion towards the horizon. 
Finally, Fig. \ref{fig:6} refers to a general non-equatorial non-circular geodesic starting from the equatorial plane with enough energy and angular momentum to escape after a few loops around the hole.
This situation exhibits both features of Fig. \ref{fig:2} and \ref{fig:3}: the deviation vector components are initially very small, then increase to high values when the orbit passes close to the hole.
During the escape far from the hole they practically stop increasing reaching asymptotic constant values.
However, the magnitude of the displacement vector is big enough so that the perturbed orbit turns out to be very different from the reference one even for small spin.

\section{Concluding remarks}

We have studied the dynamics of extended spinning bodies in the Kerr spacetime in the pole-dipole particle approximation and under the assumption that the actual motion slightly deviates from the geodesic path.
As usual, the spin parameter is assumed to be very small in order to ignore the back reaction on the spacetime geometry (see Ref. \cite{mashsingh} for a detailed discussion on the validity of this approximation).
This approach naturally leads to solve the Mathisson-Papapetrou-Dixon equations linearized in the spin variables as well as in the deviation vector, with initial conditions for geodesic motion.
Even if there exist a number of papers discussing the dynamics of extended spinning bodies in Kerr spacetime, these are but related to special situations or involve fully numerical treatments. 
The novel contribution of the present paper concerns the choice of the reference path, which is here a fully general non-equatorial and non-circular timelike geodesic, allowing to go beyond previous results limited to the very special case of deviations from an equatorial circular geodesic.

The deviations from geodesic motion has been estimated by analyzing the behavior of the components of the deviation vector as well as the magnitude of the displacement vector as functions of the proper time along the reference path.
As a general feature, even for small values of the spin parameter the deviations tend to grow with time, so that the difference between the path of a spinning particle and the reference geodesic becomes significant when measured after a time interval large enough.
We have then considered different kinds of reference geodesics, including non-circular equatorial orbits, orbits along the rotation axis and spherical orbits.
The case of polar geodesics in the Schwarzschild spacetime has been analytically solved, providing explicit expressions for radial shift and node advance of the resulting perturbed orbit.

\begin{acknowledgments}
The authors acknowledge ICRANet for support. DB thanks Dr. D. Singh for useful discussions.
\end{acknowledgments}

\appendix

\section{Electric and magnetic part of the Riemann tensor in the parallel propagated frame}
\label{app1}

The nonvanishing components of the electric part of the Riemann tensor in the parallel propagated frame are 
\begin{eqnarray}
E(U_{\rm (g)})_{\hat 1\hat 1}&=&-\frac{3Mr}{\Sigma^3K}J_3\sinh^2\beta\cosh^2\beta\cos^2\psi+\frac{Mr}{\Sigma^3}J_5\,, \nonumber\\
E(U_{\rm (g)})_{\hat 1\hat 2}&=&-\frac{3Ma\cos\theta}{\Sigma^3K}\sinh \beta\cosh \beta(J_1\cosh^2\beta-4r^2J_4)\cos\psi\,, \nonumber\\
E(U_{\rm (g)})_{\hat 1\hat 3}&=&-\frac{3Mr}{\Sigma^3K}J_3\sinh^2\beta\cosh^2\beta\cos\psi\sin\psi\,, \nonumber\\
E(U_{\rm (g)})_{\hat 2\hat 2}&=&\frac{Mr}{\Sigma^3K}[3J_3\cosh^4\beta-\cosh^2\beta(J_1-8a^2\cos^2\theta J_2)+2r^2J_5]\,, \nonumber\\
E(U_{\rm (g)})_{\hat 2\hat 3}&=&-\frac{3Ma\cos\theta}{\Sigma^3K}\sinh \beta\cosh \beta(J_1\cosh^2\beta-4r^2J_4)\sin\psi\,, \nonumber\\
E(U_{\rm (g)})_{\hat 3\hat 3}&=&\frac{3Mr}{\Sigma^3K}J_3\sinh^2\beta\cosh^2\beta\cos^2\psi-\frac{Mr}{\Sigma^3K}[3J_3\cosh^4\beta\nonumber\\
&&-4\cosh^2\beta(J_3+2a^2\cos^2\theta J_4)+r^2J_5]\,, \nonumber\\
\end{eqnarray}
where
\begin{eqnarray}
J_1&=&5r^4-10r^2a^2\cos^2\theta+a^4\cos^4\theta\,, \nonumber\\
J_2&=&3r^2-a^2\cos^2\theta\,, \nonumber\\
J_3&=&r^4-10r^2a^2\cos^2\theta+5a^4\cos^4\theta=J_1-4J_4\,, \nonumber\\
J_4&=&r^2-a^2\cos^2\theta\,, \nonumber\\
J_5&=&r^2-3a^2\cos^2\theta\,.
\end{eqnarray}

The nonvanishing components of the magnetic part of the Riemann tensor in the parallel propagated frame are 
\begin{eqnarray}
H(U_{\rm (g)})_{\hat 1\hat 1}&=&\frac{aM\cos\theta}{\Sigma^3}\left(J_2-\frac{3J_1}{K}\sinh^2\beta\cosh^2\beta\cos^2\psi\right)\,, \nonumber\\
H(U_{\rm (g)})_{\hat 1\hat 2}&=&\frac{3Mr}{\Sigma^4}\sinh \beta\cosh \beta\cos\psi\left(J_3-\frac{a^2\cos^2\theta}{K}J_1\right)\,, \nonumber\\
H(U_{\rm (g)})_{\hat 1\hat 3}&=&-\frac{3aM\cos\theta}{\Sigma^3K}\sinh^2\beta\cosh^2\beta\sin\psi\cos\psi J_1\,, \nonumber\\
H(U_{\rm (g)})_{\hat 2\hat 2}&=&\frac{3aM\cos\theta}{\Sigma^5K}\left\{(K^2-r^2a^2\cos^2\theta)J_1+\frac{K}{3}
[J_1J_2-r^2(J_1-3J_3+4\Sigma J_4)]
\right\}\,, \nonumber\\
H(U_{\rm (g)})_{\hat 2\hat 3}&=&\frac{3Mr}{\Sigma^4}\sinh \beta\cosh \beta\sin\psi\left(J_3-\frac{a^2\cos^2\theta}{K}J_1\right)\,, \nonumber\\
H(U_{\rm (g)})_{\hat 3\hat 3}&=&\frac{3aM\cos\theta}{\Sigma^3K}\left\{\left(\sinh^2\beta\cosh^2\beta\cos^2\psi-\frac{K^2-r^2a^2\cos^2\theta}{\Sigma^2}\right)J_1\right.\nonumber\\
&&\left.-\frac{2K}{3\Sigma^2}
[J_1J_2-r^2(J_1+4\Sigma J_4)]
\right\}\,. 
\end{eqnarray}


\begin{thebibliography}{00}


\bibitem{math37} 
M. Mathisson,
{Acta Phys.\ Polon.} {\bf 6}, 167 (1937).

\bibitem{papa51} 
A. Papapetrou, 
{Proc.\ R.\ Soc.\ Lond.} {\bf 209}, 248 (1951).

\bibitem{tulc59} 
W. Tulczyjew,  
{Acta Phys.\ Polon.} {\bf 18}, 393 (1959).

\bibitem{dixon64} 
W. G. Dixon, 
{Il Nuovo Cimento} {\bf 34}, 317 (1964).

\bibitem{dixon69}
W. G. Dixon, 
%``Dynamics of extended bodies in general relativity I. 
%  Momentum and angular momentum",
{Proc.\ R.\ Soc.\ Lond.\ A} {\bf 314}, 499 (1970).

\bibitem{dixon70}
W. G. Dixon, 
%``Dynamics of extended bodies in general relativity II. 
%  Moments of the charge-current vector",
{Proc.\ R.\ Soc.\ Lond.\ A} {\bf 319}, 509 (1970).

\bibitem{dixon73}
W. G. Dixon, 
{Gen.\ Rel.\ Gravit.} {\bf 4}, 199  (1973).

\bibitem{dixon74}
W. G. Dixon,  
{Philos.\ Trans.\ R.\ Soc.\ London.\ Ser.\ A} {\bf 277}, 59 (1974).

\bibitem{mohs0}
M. Mohseni, 
{Phys.\ Lett.\ B} {\bf 587}, 133  (2004).

\bibitem{maeda} 
S. Suzuki and K. Maeda, 
{Phys.\ Rev.\ D} {\bf 55}, 4848 (1997).

\bibitem{ver}
C. Verhaaren and E. W. Hirschmann, 
{Phys.\ Rev.\ D} {\bf 81}, 124034  (2010).

\bibitem{hartl1} 
M. D. Hartl,  
{Phys.\ Rev.\ D} {\bf 67}, 024005 (2003).

\bibitem{hartl2} 
M. D. Hartl, 
{Phys.\ Rev.\ D} {\bf 67}, 104023 (2003).

\bibitem{sem99}
O. Semer\'ak, 
{Mon.\ Not.\ R.\ Astron.\ Soc.} {\bf 308}, 863  (1999).

\bibitem{singh}
D. Singh, 
{Phys.\ Rev.\ D} {\bf 72}, 084033  (2005).

\bibitem{bdfg2004} 
D. Bini, F. de Felice, and A. Geralico,
{Class.\ Quantum Grav.\/} {\bf 21} 5427 (2004). 

\bibitem{bdfgj2005} 
D. Bini, F. de Felice, A. Geralico, and R. T. Jantzen, 
{Class.\ Quantum Grav.\/} {\bf 22} 2947 (2005). 
%Spin precession in the Schwarzschild spacetime: circular orbits 

\bibitem{bgj2006}
D. Bini, A. Geralico, and R. T. Jantzen, 
{Class.\ Quantum Grav.\/} {\bf 23}, 3287 (2006).
%Spin precession along circular orbits in the Kerr spacetime: the Frenet-Serret description 

\bibitem{ply}
R. Plyatsko, O. Stefanyshyn, and M. Fenyk, 
{Phys.\ Rev.\ D} {\bf 82}, 044015  (2010).

\bibitem{abrca}
M. A. Abramowicz and M. Calvani, 
{Mon.\ Not.\ R.\ Astron.\ Soc.} {\bf 189} 621, (1979). 

\bibitem{ryaba}
A. P. Ryabushko and A. A. Bakhankov, 
{Gen.\ Rel.\ Gravit.} {\bf 19}, 351  (1987).

\bibitem{svietal}
K. Svirskas, K. Pyragas, and A. Lozdiene, 
{Astrophys.\ Space\ Sci.} {\bf 149}, 39 (1988).

\bibitem{mashsingh}
B. Mashhoon and D. Singh, 
{Phys.\ Rev.\ D} {\bf 74}, 124006 (2006).

\bibitem{bgj_spindev} 
D. Bini, A. Geralico, and R. T. Jantzen,
{Gen.\ Rel.\ Gravit.} {\bf 43} 959 (2011). 

\bibitem{Marck1}
J. A. Marck, 
% Parallel tetrad on null geodesics in Kerr and Kerr-Newman space-time
{Phys.\ Lett.\ A} {\bf 97}, 140 (1983).

\bibitem{Marck2}
N. Kamran and J. A. Marck,  
% Parallel-propagated frame along the geodesics of the metrics admitting a Killing-Yano tensor
{J. \ Math.\ Phys.\ } {\bf 27}, 1589 (1986).

\bibitem{lyne}
A. G. Lyne, et al., 
{\it Science} {\bf 303}, 1153 (2004).

\bibitem{foot}
The notation used should not cause confusion. In fact, we have denoted the electric and magnetic part of the Riemann tensor as measured by an observer $U$ by $E(U)$ and $H(U)$, respectively.
The electric and magnetic part of the Killing-Yano tensor are instead denoted by ${\mathcal E}(U)$ and ${\mathcal B}(U)$.
Similarly, projector operator orthogonal to the timelike direction $U$ is denoted by $P(U)$ whereas ${\mathcal P}(U)$ is the Poynting vector associated with the Killing-Yano tensor.

\bibitem{punslychmash}
C. Chicone, B. Mashhoon, and P. Punsly, 
{Phys.\ Lett.\ A} {\bf 343}, 1 (2005).



\end{thebibliography}
\end{document}